\documentclass[aps,pre,twocolumn,showpacs,preprintnumbers,amsmath,amssymb,epsfig,reprint]{revtex4-2}

\usepackage{epsfig,amssymb}
\usepackage{color}
\usepackage{xcolor}
\usepackage{graphicx}
\usepackage{comment}
\usepackage{ulem}
\usepackage{datetime}

\usepackage{float}

\newcommand{\mycomment}[1]{}

\usepackage{hyperref}

\begin{document}

\draft

\title {Common Noise-Induced Group-Level Synchronization \linebreak Between Uncoupled Groups of Oscillators}

\author{Tae-Wook Ko}
\email{twko@nims.re.kr}
\affiliation{National Institute for Mathematical Sciences, Daejeon 34047, Republic of Korea}
\date{\today}

\begin{abstract}
We investigate group-level synchronization between oscillator groups induced by common noise in the absence of inter-group coupling. Each group receives a common noise shared by all its oscillators and independent local noise inputs to individual oscillators. The same common noise is applied to all groups.
The system is studied with both identical and nonidentical oscillators, and with and without intra-group coupling. In the nonidentical case, natural frequencies are drawn from  the same distribution for both groups, making them  statistically equivalent.
Through numerical simulations of this system, we find that the degree of synchronization within each group, measured by the absolute value of a complex Kuramoto order parameter, typically shows significant temporal fluctuations.
Importantly, the complex order parameters representing the collective oscillations of the groups synchronize when the groups are driven by the same common noise.
By deriving a phase density evolution mapping, we          analytically explain how this group-level synchronization   is achieved in the absence of intra-group coupling.
\end{abstract}

\maketitle

\section{Introduction}
Synchronization of oscillations represents a fundamental organizing principle in complex systems ranging from neural circuits to chemical reactions and physical systems \cite{winfree2001,pikovsky_book,sync,kura,strogatz2000,kura_review,ermentrout2001}. 
While much research has focused on synchronization emerging through direct coupling between oscillators, it has also been shown that even uncoupled identical oscillators can achieve full synchronization when subjected to common noise input, regardless of their initial conditions \cite{pikovskii1984,goldobin2004,teramae2004,galan2006,galan2007a,galan2007b,marella2008,abouzeid2009}. 
As noted in Refs. \cite{goldobin2005,goldobin2006,galan2007a}, there exists a correspondence between this noise-induced synchronization of identical oscillators and the phenomenon of neuronal reliability, where a single neuron generates the same spike sequence when the same fluctuating input is repeatedly applied, despite different initial states \cite{mainen1995}.
While common noise input alone leads to full synchronization in systems of uncoupled identical oscillators, when oscillators receive independent local noise inputs in addition to the common noise input, they exhibit correlated but not fully synchronous oscillations \cite{goldobin2005,nakao2007,abouzeid2011}. 
In this case, the phase differences between oscillators follow a broad distribution peaked at zero, indicating phase difference fluctuation around zero \cite{goldobin2005,nakao2007,abouzeid2011}.
It was shown in Ref. \cite{nagai2010} that in coupled nonidentical oscillators receiving common noise the critical coupling strength for the transition from the incoherent state to the synchronous state is reduced by the noise. It was also shown that uncoupled nonidentical oscillators can be synchronized by common noise input alone and the degree of synchrony can be enhanced by increasing the noise strength \cite{lai2013}. 

Here, we focus on collective dynamics of oscillator groups rather than individual-level dynamics when oscillators receive both local and common noise. More specifically, we analyze how group-level behaviors synchronize across different groups when the groups are under the influence of the same common noise input, despite having no direct coupling between groups.
This group-level perspective is particularly relevant for understanding complex biological and physical systems where group-level dynamics are key to determining system behavior or are more readily observable than individual-level dynamics.
Moreover, different groups influenced by the same external fluctuations are frequently encountered in natural systems. An effect known as the Moran effect \cite{moran1953,hudson1999,hansen2020} has been observed where ecological populations distributed over spatially separated habitats exhibit synchronized fluctuations when exposed to common climatic influences. 
In regions where the well-mixed assumption does not hold, the system is better represented by dividing the region into multiple well-mixed sub-regions,  each of which is described by a single oscillator. 
Spatially separated regions of this kind can then be viewed as uncoupled groups of oscillators under localized environmental variations and broader climatic influences shared across all regions. 
This ecological configuration of multiple oscillator groups under shared external fluctuations corresponds well to the framework we analyze in this paper.
Neuronal populations in distinct brain regions may synchronize their collective activity when responding to the  same sensory input. Similarly, when different observers view the same time-varying sensory input such as a movie, their brain waves, which correspond to the collective neural activities, exhibit inter-subject synchronization \cite{hasson2004,hasson2009,denworth2023}.
These natural systems are examples showing common external inputs can coordinate group-level behaviors despite the absence of direct coupling, a phenomenon we explore in this work.

In Refs. \cite{lin2009a,lin2009b}, the authors examined pulse-coupled networks of nonidentical phase oscillators. They demonstrated that the pooled response, the total output from a subpopulation of neurons, is reliably reproduced when the same fluctuating input is repeatedly applied, regardless of initial condition of the network. This reliability occurs even when individual oscillators do not exhibit reliable responses. They also observed that independent local noise inputs to individual oscillators do not significantly affect the reliability.
This pooled response reliability directly relates to our focus on group-level behavior. In fact, it can be conceptually translated into what we term group-level synchronization between uncoupled identical groups of oscillators receiving common input, just as individual-level reliability corresponds to noise-induced synchronization between single oscillators.
While pulse coupling effectively models discrete neuronal interactions, 
continuous coupling has been widely adopted for modeling various dynamical systems \cite{winfree2001,pikovsky_book,sync,kura,strogatz2000,kura_review,ermentrout2001}. 
In this study, by examining systems with continuous coupling, we complement existing pulse-coupling research while investigating how local noise and common noise affect collective behavior, and specifically, how sharing common noise among groups induces group-level synchronization.
In particular, previous work \cite{kawamura2008} considered two isolated populations of coupled identical oscillators, each receiving local noise and exhibiting coherent collective oscillations. They showed that introducing common noise to both populations synchronizes these collective oscillations between the two populations. 
We extend this framework to cases of (non)identical oscillators with and without    intra-group coupling, where coherent collective oscillations may or may not pre-exist. We note that the group-level synchronization in Ref. \cite{kawamura2008} can be understood as noise-induced synchronization of uncoupled limit-cycle oscillators, where each group acts as a single limit-cycle oscillator. 
 In contrast, our cases include situations where each group cannot be described as a single oscillator and thus falls outside the framework of noise-induced synchronization of limit-cycle oscillators, requiring a different approach.

In this paper, we investigate group-level synchronization between oscillator groups driven by both independent local noise and shared common noise, in the absence of inter-group coupling.
We study both identical and nonidentical oscillators, with and without intra-group coupling. Through numerical simulations, we demonstrate that the degree of synchronization within a single group exhibits significant temporal fluctuations, and that group-level synchronization occurs when they share the same common noise. Using a phase density evolution mapping, we provide analytical explanations for this group-level synchronization in the absence of intra-group coupling.

\section{Model}
We consider two groups of limit-cycle oscillators receiving additive noise, 
where oscillators within each group can be uncoupled or all-to-all equally coupled, but the two groups remain uncoupled from each other.
The noise inputs to the oscillators are partially correlated due to the shared common noise input. This allows us to investigate how oscillator heterogeneity and intra-group coupling affect the collective response to common external fluctuating input.
With the assumption of weak coupling and weak additive white noise, the system can be reduced to the phase-only system described by the following stochastic differential equations (SDEs) in the It\^{o} sense \cite{abouzeid2011,teramae2009,goldobin2010}.
\begin{subequations}
\begin{align}
	\label{eq:model_1a}
	d\theta_{gi} &= \Bigl [ \omega_{gi} +\frac{K}{N} \sum_{j=1}^{N}H(\theta_{gj}-\theta_{gi}) \nonumber \\
	&\quad+ \frac{\sigma^2}{2}\Delta(\theta_{gi}) \Delta'(\theta_{gi})\Bigr ]dt
	+ \sigma\Delta(\theta_{gi})dW_{gi},\\
	\label{eq:model_1b}
	dW_{gi} &= \sqrt{c_{in}}\,d\xi_{gc} + \sqrt{1-c_{in}}\,d\xi_{gi},\\
& ~g = 1, 2, \nonumber \\
	& ~i = 1, 2, ..., N, ~~c_{in} \in [0,1], \nonumber
\end{align}
\label{eq:model}
\end{subequations}
where $gi$ denotes the $i$th oscillator of group $g$, with $\theta_{gi}$ representing its phase at time $t$, $\omega_{gi}$ its natural frequency, and $N$ the total number of oscillators in each group. The natural frequencies $\omega_{gi}$ are either identical, where $\omega_{gi}=\omega_0$ for all $i$ and $g$, or heterogeneous, drawn from a Gaussian distribution $p(\omega)$ with mean $\omega_0$ and standard deviation $\sigma_\omega$.

The second term in the bracket of Eq.~(\ref{eq:model_1a}) represents the intra-group coupling. The coupling strength $K(\geq 0)$ and coupling function $H(\theta)$ characterize the interaction between oscillators, where $H(\theta)$ is $2\pi$-periodic in the phase difference $\theta$ between two oscillators.
Here, we use $H(\theta)=\sin\theta$, 
which promotes synchronization for positive coupling strength and has been widely adopted in studies of coupled oscillators because it facilitates analytical treatment \cite{kura,strogatz2000,kura_review}.

The third term within the bracket is the additional term in the deterministic part 
arising from the phase reduction of the stochastic system, which is absent in the phase-reduced model of the corresponding deterministic system.
The parameter $\sigma$ represents the noise strength, while $\Delta(\theta)$ is the phase resetting curve (PRC) describing how the effect of the input on the phase advance or delay depends on the current phase \cite{canavier2006,smeal2010}.
Unless otherwise stated, we primarily study the model with a type II PRC $\Delta(\theta) = -\sin\theta$, which can produce either phase advances or delays depending on when the perturbation occurs. For comparison, we also present results in Appendix B with a type I PRC $\Delta(\theta) = 1-\cos\theta$, which produces exclusively phase advances. Both PRCs give qualitatively similar results.

The term outside the bracket of Eq.~(\ref{eq:model_1a}) is the stochastic term, where $dW_{gi}$ is the noise input  to the oscillator $gi$ given by Eq.~(\ref{eq:model_1b}), and the effect of this input on the phase is given by the product of  $dW_{gi}$ and  the phase resetting curve $\Delta(\theta)$.
In Eq.~(\ref{eq:model_1b}), $\xi_{gc}$ and $\xi_{gi}$ for $g=1,2$ and $i=1,2,...,N$ are independent Wiener processes, and their increments $d\xi_{gc}$ and $d\xi_{gi}$ are mutually independent except for the possible case where $d\xi_{1c} = d\xi_{2c}$. 
The increment $d\xi_{gc}$ is the common component shared by all oscillators within group $g$, while $d\xi_{gi}$ is the independent local component of oscillator $gi$.  The two groups may receive either the same common component ($d\xi_{1c} = d\xi_{2c}$) or different common components ($d\xi_{1c} \neq d\xi_{2c}$), and the former case is the key condition for group-level synchronization.
The term $\sigma\sqrt{c_{in}}d\xi_{gc}$ acts as the common noise input to the oscillators of group $g$ with $c_{in}$   being the correlation coefficient among the noise inputs $dW_{gi}$ within the group. 
By adjusting $c_{in}$, we can explore how different levels of noise correlation within each group affect the behavior of the system and the group-level synchronization between the two groups.

\begin{figure*}
\centering
\epsfig{figure=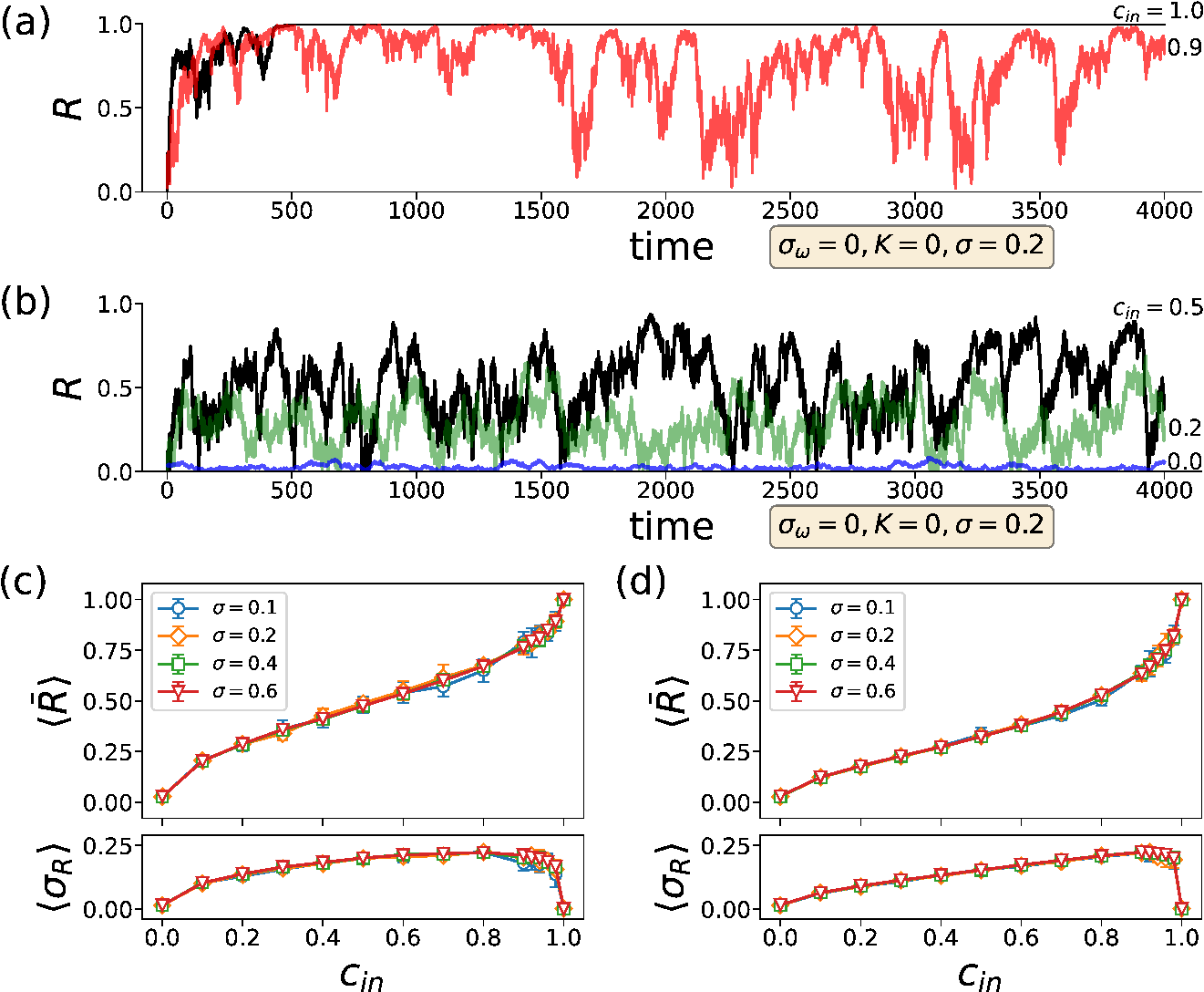, width= 13.8cm}
	\caption{Order parameter $R$ of a group of uncoupled identical oscillators ($\sigma_{\omega} = 0$, $K=0$): 
	(a) $R(t)$ for $c_{in}=1.0$ and $c_{in}=0.9$ (b) $R(t)$ for $c_{in}=0.5$, $c_{in}=0.2$, and $c_{in}=0.0$. Simulations of (a) and (b) are performed with $\sigma=0.2$ and $\Delta(\theta) = -\sin\theta$. 
	(c) $\langle \bar R \rangle$ and $\langle \sigma_R \rangle$ as a function of $c_{in}$ with $\Delta(\theta) = -\sin\theta$. (d) $\langle \bar R \rangle$ and $\langle \sigma_R \rangle$ as a function of $c_{in}$ with $\Delta(\theta) = 1-\cos\theta$. 
Initial values of phases are randomly selected from $[0, 2\pi)$.  
Different random seeds are used for generating the noise inputs and the initial phases for different simulations. 
For the details, see the text.
	}
\label{fig:ucpld_w_one}
\end{figure*}

To characterize the dynamics of the oscillator population, we measure the degree of synchronization of group $g$ using the complex Kuramoto order parameter defined as: 
\begin{align}
	Z_g=R_g(t) {\rm e}^{\mathrm{i}\Theta_g(t)} &= \frac{1}{N}\sum_{j=1}^{N} {\rm e}^{\mathrm{i}\theta_{gj}(t)},
\label{eq:op}
\end{align}
where $R_g$ is the Kuramoto order parameter of group $g$ and $\Theta_g$ is the corresponding phase.  
Let $\mathrm{i}=\sqrt{-1}$ denote the imaginary unit.
The Kuramoto order parameter $R_g$, which represents the degree of synchronization, ranges from 0 to 1 corresponding to incoherence and in-phase synchronization, respectively.  
The complex Kuramoto order parameter can be interpreted as a collective oscillation of the group.
For simplicity in subsequent discussions, we often omit the group index $g$ and denote quantities like $R_g$ simply as $R$ when the group context is clear.

\section{\label{sim_results}Simulation Results}
\subsection{Numerical methods}
We conduct numerical simulations of Eq.~(\ref{eq:model}) using the Euler-Maruyama method, a numerical approach well-suited for solving SDEs in the It\^{o} sense \cite{higham2001}.
For the simulations, we set $\omega_0 = 2\pi$, $N=1000$ and a time step of $\delta t = 0.01$.  The specific choice of $\omega_0$ does not change the results qualitatively.

We calculate the time average of quantities of interest over a time interval $[t = T_1, t = T_2]$, excluding the initial transient behavior. For most cases without intra-group coupling, we set $T_1 = 1000$ and $T_2 = 4000$. However, for cases with the noise strength $\sigma=0.1$, we use a longer time interval of $T_1 = 4000$ and $T_2 = 12000$ to ensure sufficient data are collected for accurate time averaging due to the longer transient dynamics at this noise level. Additionally, for cases with intra-group coupling that can exhibit irregular group-level desynchronization events, we also use $T_1 = 4000$ and $T_2 = 12000$ to allow reliable time averaging.

\subsection{\label{sbsec:ucpld_w}
{Two groups of uncoupled identical oscillators}}
Let us first consider the case of two groups of uncoupled identical oscillators
($\sigma_\omega=0$ and $K = 0$). 

Figures~\ref{fig:ucpld_w_one}(a) and (b) show the time evolutions of the order parameter $R$ of a single group for different values of $c_{in}$. 
Initially, the system is in the uniformly incoherent state where the phases of the oscillators are selected from $[0, 2\pi)$.
The initial condition does not affect the long-term behavior of the system.
For the cases with perfectly correlated noises ($c_{in} = 1$), that is, 
all oscillators receiving the same noise input within a group, the oscillators starting from different initial phases become in-phase synchronized ($R(t):  0 \rightarrow 1$) (Fig.~\ref{fig:ucpld_w_one}(a)). 
On the contrary, with near-zero correlation between the noises ($c_{in} \approx 0$), the oscillators behave incoherently ($R(t) \approx 0$) (Fig.~\ref{fig:ucpld_w_one}(b)).

For intermediate levels of correlation ($c_{in} = 0.9$, $0.5$, and $0.2$), the order parameter $R$ exhibits behavior distinct from the extreme cases ($c_{in} \approx 0$ and $c_{in} = 1$). While $R$ remains below 1, it fluctuates significantly, ranging from near zero to high values (Figs.~\ref{fig:ucpld_w_one}(a) and (b)). 
This intermediate behavior reflects nontrivial interplay between noise correlation and system dynamics.

To show how the order parameter and its fluctuations vary with the correlation $c_{in}$, we measure $\langle \bar R \rangle$ and $\langle \sigma_R \rangle$, where $\bar R$ is the time average of $R$ over $[T_1,T_2]$, $\langle \cdot \rangle$ is the average over $10$ simulations, and $\sigma_R$ is the standard deviation of $R$ over $[T_1,T_2]$.                                
Figures~\ref{fig:ucpld_w_one}(c) with $\Delta(\theta) = -\sin\theta$ and (d) with $\Delta(\theta) = 1-\cos\theta$ illustrate how $\langle \bar R \rangle$ and $\langle \sigma_R \rangle$ vary with the correlation coefficient $c_{in}$. 
As observed in Figs.~\ref{fig:ucpld_w_one}(a) and (b), $\langle \bar{R} \rangle$
monotonically increases from $0$ to $1$ as $c_{in}$ increases from $0$ to $1$, while
the curves for $\langle \sigma_R \rangle$ are concave unimodal functions that
first increase from near zero and then decrease to zero.
For a wide range of $c_{in}$ values, $\sigma_R$ is comparable to $\bar R$.
This indicates that the fluctuations in the degree of synchronization
are as large as the mean degree of synchronization itself.
Note that the curves collapse regardless of the noise strength $\sigma$. 
As the noise strength $\sigma$ increases for $0<c_{in} < 1$, the order parameter $R$ fluctuates more frequently, but both the time-averaged value $\bar R$ and the standard deviation $\sigma_R$ remain statistically unchanged for fixed $c_{in}$.

Extensive studies have examined equivalent systems of two uncoupled identical oscillators under common noise \cite{teramae2004,galan2006,galan2007a,galan2007b,marella2008,abouzeid2009,nakao2007,abouzeid2011}. These studies established that under fully correlated noise, two uncoupled oscillators synchronize in-phase, which implies that a group of oscillators receiving the same common noise also synchronize and the order parameter eventually reaches one \cite{teramae2004,galan2006,galan2007a,galan2007b,marella2008,abouzeid2009}. Conversely, with completely uncorrelated noise, oscillators behave incoherently, explaining the incoherent group behavior we observe in this regime.
For intermediate correlation levels, previous work showed that pairwise phase differences follow a broad distribution centered at zero, suggesting fluctuations around the synchronized state \cite{nakao2007,abouzeid2011}.
However, from this result alone, one cannot predict whether the order parameter fluctuates or remains constant. If all pairwise phase differences simply maintained the same distribution over time, the order parameter would stay constant. Our findings show that the order parameter actually exhibits significant temporal fluctuations.

\begin{figure*}
\centering
\epsfig{figure=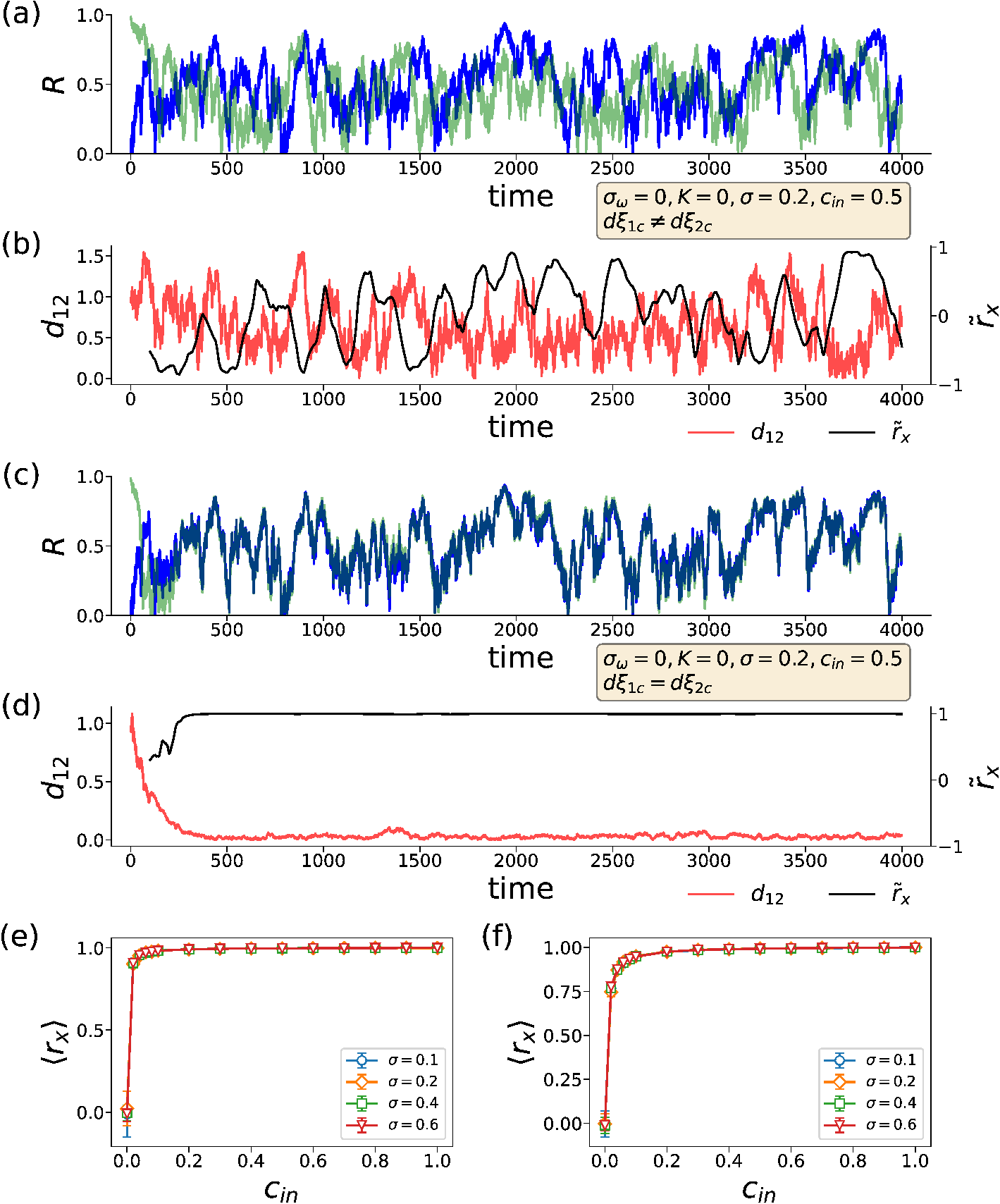, width= 13.8cm}
	\caption{Two groups of uncoupled identical oscillators ($\sigma_{\omega} = 0$, $K=0$): 
	(a) With different common noise inputs to the groups ($d\xi_{1c} \neq d\xi_{2c}$): asynchronous behaviors of $R_1(t)$ and $R_2(t)$ for $c_{in}=0.5$. 
	(b) $d_{12}(t)$ and $\tilde r_x(t)$ for (a). 
	(c) With the same common noise input to the groups ($d\xi_{1c} = d\xi_{2c}$): synchronized behaviors of $R_1(t)$ and $R_2(t)$ for $c_{in}=0.5$. 
	(d) $d_{12}(t)$ and $\tilde r_x(t)$ for (c). 
	Simulations of (a)-(d) are performed with $\sigma=0.2$ and $\Delta(\theta) = -\sin\theta$. 
	(e) $\langle r_x \rangle$ as a function of $c_{in}$ with $\Delta(\theta) = -\sin\theta$. (f) $\langle r_x \rangle$ as a function of $c_{in}$ with $\Delta(\theta) = 1-\cos\theta$. 
	In (a) and (c), initial values of phases are randomly selected from $[0, 2\pi)$ for group 1 (red curve) and from $[0, 0.2\pi)$ for group 2 (green curve).  
Different random seeds are used for generating the noise inputs and the initial phases for different simulations. 
For the details, see the text.
	}
\label{fig:ucpld_w_two}
\end{figure*}

Next, we investigate the effect of applying identical common noise to both groups ($d\xi_{1c} = d\xi_{2c}$). To highlight this effect, we initialize the groups with contrasting states: a near-incoherent state and a near-synchronous state, respectively.
As a baseline comparison, Fig.~\ref{fig:ucpld_w_two}(a) shows the time series of $R$ for both groups receiving different common noises ($d\xi_{1c} \neq d\xi_{2c}$) at $c_{in}=0.5$.
As expected, the time evolutions of $R(t)$ are different. To quantify the dissimilarity between the collective oscillations of the two groups, we measure the absolute value of the difference between the two complex order parameters:
\begin{equation}
	d_{12}(t)\equiv \left |Z_1 (t) -Z_2 (t) \right |. 
    \label{d12}
\end{equation}
For the case of Fig.~\ref{fig:ucpld_w_two}(a), $d_{12}$ fluctuates widely over the simulation time (Fig.~\ref{fig:ucpld_w_two}(b) red line).
Note that when the $R$ values of the groups are small, this measure can give small values regardless of the synchrony between the two groups. 
To overcome this ambiguity, we additionally measure the sliding window Pearson correlation coefficient (SWPCC) $\tilde {r}_x(t)$ between ${Z}_{1x}$ and ${Z}_{2x}$ over sliding time windows $[t-w,t]$ with window size $w=100$ and sliding step size $s=0.2$, where ${Z}_{gx} = R_g \cos\Theta_g$ (see Appendix A). 
When the complex order parameters of the two groups are linearly independent, $\tilde {r}_x(t)$ can be close to zero or fluctuate within the range $[-1,1]$.
In contrast, when they are highly correlated, $\tilde {r}_x(t)$ approaches $1$. 
In Fig.~\ref{fig:ucpld_w_two}(b), which shows asynchronous behaviors of $Z_1$ and $Z_2$, $\tilde {r}_x$ denoted by a black line fluctuates over $[-1, 1]$.

In general, a Pearson correlation coefficient of $1$ does not necessarily indicate identical time evolution of two signals. It only guarantees linear correlation. However, in our system, sustained high correlation $\tilde {r}_x(t) \approx 1$ across multiple time windows indicates that the phase distributions of the two groups must be similar and  evolving similarly. If the phase distributions were significantly different, $\tilde {r}_x(t)$
would not remain close to $1$ in subsequent time windows.
Note that as shown in Appendix A, $\tilde {r}_y(t) \approx \tilde {r}_x(t)$, and therefore $\tilde {r}_x \approx 1$
alone is sufficient to conclude that $Z_{1}$ and $Z_{2}$ are synchronized in the corresponding time window.

When identical common noise is applied ($d\xi_{1c} = d\xi_{2c}$, $c_{in}=0.5$), $R$ values of both groups come to fluctuate identically after initial transients (Fig.~\ref{fig:ucpld_w_two}(c)).
Figure~\ref{fig:ucpld_w_two}(d) shows that $d_{12}(t) \rightarrow 0$ and $\tilde {r}_x \rightarrow 1$  over time and this clearly indicates that the shared common noise induces the synchronization of $Z_{1}$ and $Z_{2}$. 
To investigate how noise correlation $c_{in}$ affects the group-level synchronization between the two groups, we measure the Pearson correlation coefficient $r_x$ between $Z_{1x}$ and $Z_{2x}$ over the time interval $[T_1, T_2]$ for different values of noise strength $\sigma$. 
Like $\tilde r_x$, $r_x \approx 1$ alone is sufficient to imply that $Z_1$ and $Z_2$ are synchronized.
Figures~\ref{fig:ucpld_w_two}(e) with $\Delta(\theta)=-\sin\theta$ and (f) with $\Delta(\theta)=1-\cos\theta$ show $\langle r_x \rangle$ as a function of $c_{in}$.
Remarkably, the two groups maintain strong group-level synchronization ($\langle r_x \rangle \approx 1$) even at low noise correlations, persisting down to $c_{in} \approx 0.1$.

These results show that for uncoupled identical oscillators, the common noise induces significant temporal fluctuations in the order parameter for intermediate noise correlations, while the sharing of the common noise across groups induces group-level synchronization robustly even at low noise correlations.

\subsection{\label{subsec:ucpld_wi}Two groups of uncoupled nonidentical oscillators}
\begin{figure*}
\centering
\epsfig{figure=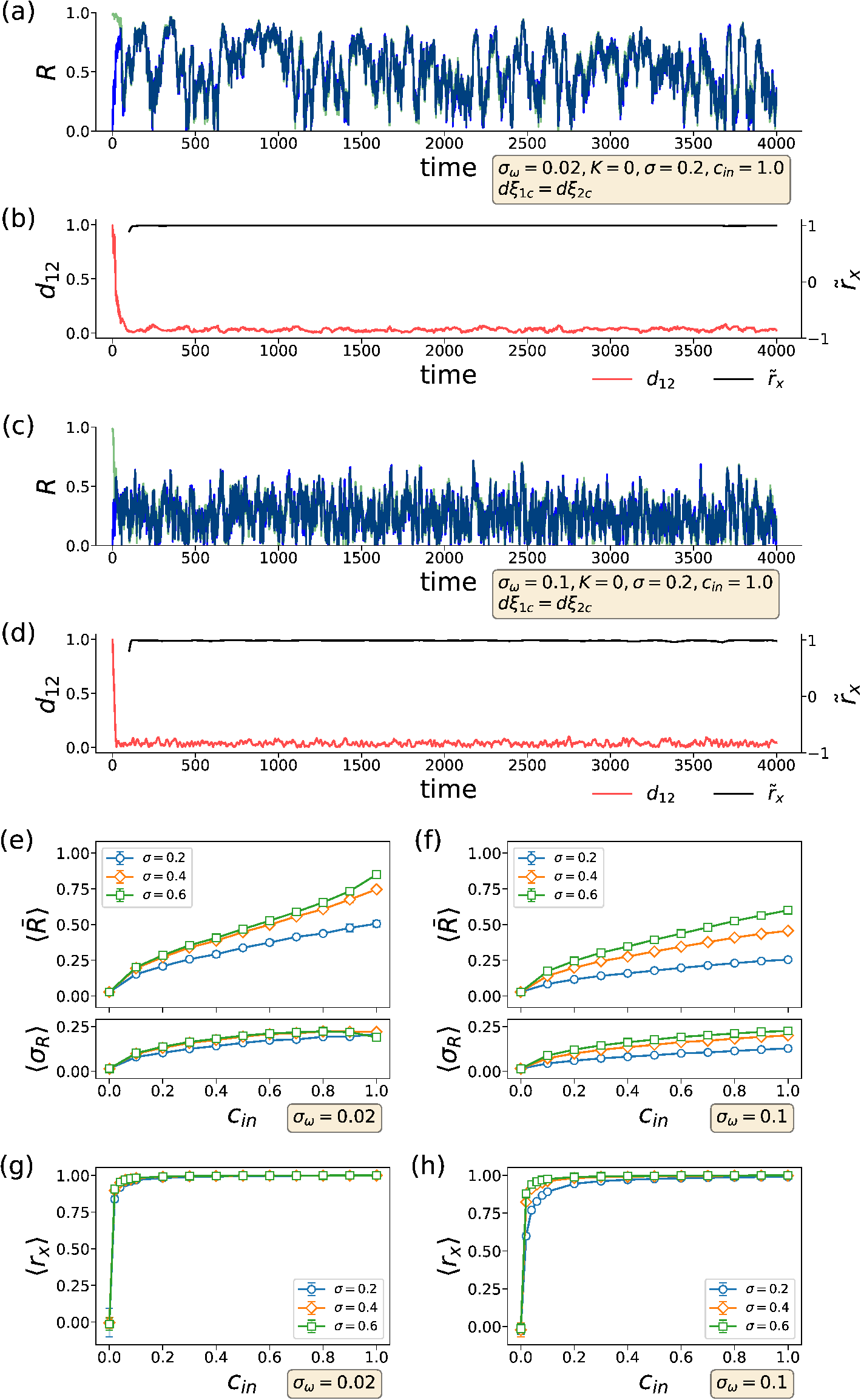, width= 13.8cm}
	\caption{Two groups of uncoupled nonidentical oscillators ($\sigma_{\omega} \neq 0$, $K=0$) with the same common noise input to the groups ($d\xi_{1c} = d\xi_{2c}$): 
	(a) Synchronized behaviors of $R_1(t)$ and $R_2(t)$ for $\sigma_{\omega} = 0.02$, $\sigma = 0.2$, and $c_{in} = 1$.	
	(b) $d_{12}(t)$ and $\tilde r_x(t)$ for (a). 
	(c) Synchronized behaviors of $R_1(t)$ and $R_2(t)$ for $\sigma_{\omega} = 0.1$, $\sigma = 0.2$, and $c_{in} = 1$.	
	(d) $d_{12}(t)$ and $\tilde r_x(t)$ for (c). 
	$\langle \bar R \rangle$ and $\langle \sigma_R \rangle$ as functions of $c_{in}$ for (e) $\sigma_{\omega} = 0.02$ and (f) $\sigma_{\omega} = 0.1$. 
	$\langle r_x \rangle$ as functions of $c_{in}$ for (g) $\sigma_{\omega} = 0.02$ and (h) $\sigma_{\omega} = 0.1$. 
	$\Delta(\theta) = -\sin\theta$ for all the figures.
	Initial conditions are as in Fig.~2.
}
\label{fig:ucpld_wi}
\end{figure*}

\begin{figure*}
\centering
\epsfig{figure=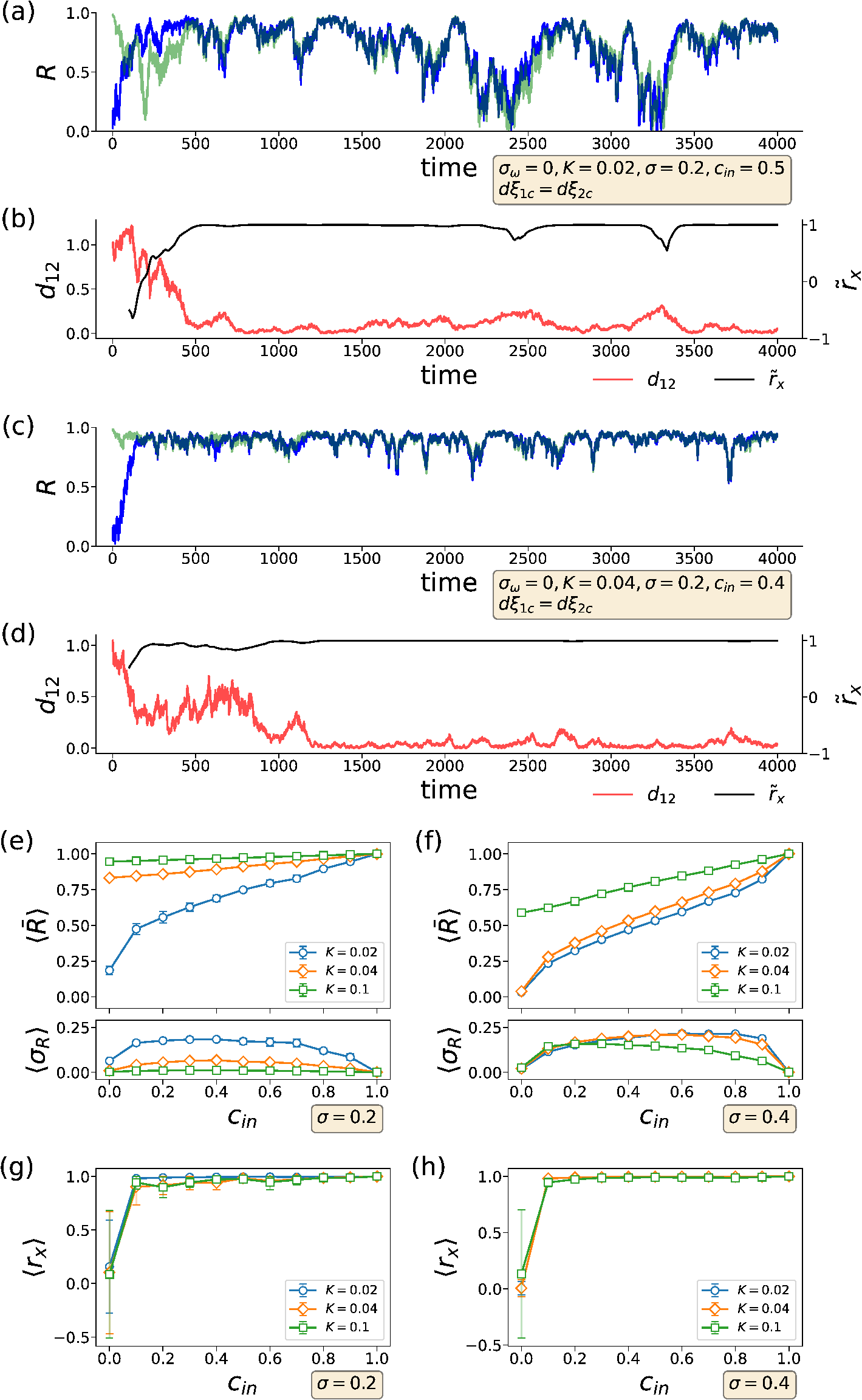, width= 13.8cm}
\caption{Two groups of coupled identical oscillators ($\sigma_{\omega} = 0$, $K>0$) with the same common noise input to the groups ($d\xi_{1c} = d\xi_{2c}$) : 
	(a) Synchronized behaviors of $R_1(t)$ and $R_2(t)$ with intermittent desynchrony for $K=0.02$, $\sigma = 0.2$, and $c_{in} = 0.5$.	
	(b) $d_{12}(t)$ and $\tilde r_x(t)$ for (a). 
	(c) Synchronized behaviors of $R_1(t)$ and $R_2(t)$ for $K=0.04$, $\sigma = 0.2$, and $c_{in} = 0.4$.	
	(d) $d_{12}(t)$ and $\tilde r_x(t)$ for (c). 
	$\langle \bar R \rangle$ and $\langle \sigma_R \rangle$ as functions of $c_{in}$ for (e) $\sigma=0.2$ and (f) $\sigma=0.4$. 
	$\langle r_x \rangle$ as functions of $c_{in}$ for (g) $\sigma=0.2$ and (h) $\sigma=0.4$. 
	$\Delta(\theta) = -\sin\theta$ for all the figures.
	Initial conditions are as in Fig.~2.
	}
\label{fig:cpld_w}
\end{figure*}

We now examine the case of two groups of uncoupled nonidentical oscillators ($\sigma_\omega \neq 0$ and $K = 0$).
Without noise ($\sigma = 0$), the oscillators of a single group eventually show incoherent behavior ($R_g \approx 0$) regardless of the initial conditions.

	Figure~\ref{fig:ucpld_wi} displays simulation results of two groups of uncoupled nonidentical oscillators with the same common noise applied to the two groups ($d\xi_{1c} = d\xi_{2c}$).
	With the application of perfectly correlated noise ($c_{in} = 1$), the nonidentical oscillators of a single group behave differently from identical oscillators. 
The oscillators, whose natural frequencies are drawn from a Gaussian 
distribution $p(\omega)$ with mean $\omega_0$ and standard deviation
$\sigma_\omega=0.02$, do not synchronize fully. Instead, they show temporal
fluctuations between high synchronization levels ($R_g \approx 1$) and low
synchronization levels ($R_g \approx 0$) (Fig.~\ref{fig:ucpld_wi}(a)).
	The two groups respectively starting from near-incoherence and near-synchrony  achieve group-level synchronization as in the cases of uncoupled identical oscillators (Figs.~\ref{fig:ucpld_wi}(a) and (b)). 

	As we increase the heterogeneity of the natural frequencies by increasing $\sigma_\omega$ to $0.1$ in Fig.~\ref{fig:ucpld_wi}(c) with other parameters fixed, the degree of synchronization $R_g$ decreases and fluctuates faster. We also observe the same group-level synchronization of the two groups with the increased $\sigma_\omega$ (Figs.~\ref{fig:ucpld_wi}(c) and (d)).

	Figures~\ref{fig:ucpld_wi}(e)-(h) summarize the simulation results with different levels of 
	natural frequency heterogeneity $\sigma_\omega$, noise strength $\sigma$, and noise correlation $c_{in}$. 
	For fixed values of $\sigma_\omega$ and $\sigma$, as $c_{in}$ increases from $0$ to $1$, $\langle \bar R\rangle$ and $\langle \sigma_R \rangle$ monotonically increase from $0$ to finite values less than $1$ (Figs.~\ref{fig:ucpld_wi}(e) and (f)), in contrast to the cases of uncoupled identical oscillators where $\langle \sigma_R \rangle$ shows a concave unimodal behavior. Therefore, unlike the cases of uncoupled identical oscillators, full synchronization ($R_g=1$) within a group cannot be achieved even with perfect correlation ($c_{in}=1$) and large fluctuations remain for higher values of $c_{in}$.

	Increasing noise strength from $\sigma=0.2$ to $\sigma=0.6$ increases $\langle \bar R\rangle$ and $\langle \sigma_R \rangle$ for fixed values of $\sigma_\omega$ and $c_{in}$. This indicates that stronger noise enhances intra-group synchronization while also increasing its fluctuations.
    As expected, increasing the natural frequency heterogeneity $\sigma_\omega$ decreases intra-group synchronization for fixed values of $\sigma$ and $c_{in}$.

As shown in Fig.~\ref{fig:ucpld_wi}(g), for small natural frequency heterogeneity $\sigma_\omega=0.02$, we observe strong group-level synchronization ($\langle r_x \rangle \approx 1$) even at noise correlations as low as $c_{in} \approx 0.1$. There is no significant difference for the three levels of noise strengths $\sigma=0.2$, $\sigma=0.4$, and $\sigma=0.6$. In contrast, with larger $\sigma_\omega=0.1$, the range for strong group-level synchronization is reduced to $c_{in} \gtrsim 0.3$ (Fig.~\ref{fig:ucpld_wi}(h)). The decrease in $\langle r_x \rangle$ is more pronounced for smaller values of noise strength $\sigma$.

For uncoupled nonidentical oscillators, the common noise induces intra-group synchronization and significant fluctuations in the degree of synchronization that persist even for higher values of $c_{in}$. In contrast to the cases of uncoupled identical oscillators, full synchronization cannot be achieved due to frequency heterogeneity. Nevertheless, the sharing of the common noise across groups still induces group-level synchronization.

\subsection{Two groups of coupled identical oscillators}
Next, we investigate the case of two groups of coupled identical oscillators
($\sigma_\omega=0$ and $K > 0$). 
Here, we explore how the common noise input affects the collective behavior of the groups and the group-level synchronization in the presence of intra-group coupling. 
Without noise, each group becomes in-phase synchronized for any $K > 0$. 

Figures~\ref{fig:cpld_w}(a)-(d) display the collective behaviors of the two groups and group-level synchronization. 
As in the cases of two groups of uncoupled identical oscillators, the groups show fluctuating order parameters for the intermediate values of $c_{in}$ (Figs.~\ref{fig:cpld_w}(a) and (c)) and group-level synchronization (Figs.~\ref{fig:cpld_w}(b) and (d)). 
Compared to the cases of uncoupled identical oscillators, it takes more time to achieve group-level synchronization in this system (Figs.~\ref{fig:cpld_w}(a)-(d)), and it occasionally experiences brief periods of slight group-level desynchronization, as evidenced by small dips in ${\tilde r}_x$, before quickly recovering group-level synchronization (Fig.~\ref{fig:cpld_w}(b)).

	In Figs.~\ref{fig:cpld_w}(e)-(h), we explore how $K$, $c_{in}$, and $\sigma$ affect collective behaviors and group-level synchronization.
	For fixed values of $\sigma$ and $K$, the degree of synchronization $\langle \bar R\rangle$ monotonically increases from a (non)zero value to $1$ as $c_{in}$ increases from $0$ to $1$ (Figs.~\ref{fig:cpld_w}(e) and (f)), while the curves for $\langle \sigma_R \rangle$ are concave unimodal functions that first increase from near zero and then decrease to zero.
	The nonzero value of $\langle \bar R\rangle$ with $c_{in}=0$ is due to the coupling, which is strong enough to overcome the desynchronizing effect of noise. 
    For fixed values of $\sigma$ and $c_{in}$, $\langle \bar R\rangle$ increases while $\langle \sigma_R \rangle$ decreases as we increase $K$. 
	Increasing $\sigma$ decreases $\langle \bar R\rangle$ for fixed values of $K$ and $c_{in}$.

As in the cases of uncoupled oscillators, we observe strong group-level synchronization in the presence of intra-group coupling for a wide range of $c_{in}$ (Figs.~\ref{fig:cpld_w}(g) and (h)).  
For noise strength $\sigma=0.2$ and weak coupling strength $K=0.02$, strong group-level synchronization ($\langle r_x \rangle \approx 1$) is observed for $c_{in} \gtrsim 0.1$.
With larger coupling strengths $K=0.04$ and $K=0.1$, the ranges for strong group-level synchronization are reduced to higher $c_{in}$ values.
Larger coupling strength helps intra-group synchronization and reduces its fluctuations but can hinder group-level synchronization.
When we increase the noise strength to $\sigma=0.4$ in Fig.~\ref{fig:cpld_w}(h), strong group-level synchronization is achieved with $c_{in}\gtrsim 0.1$ for all the same coupling strengths considered in Fig.~\ref{fig:cpld_w}(g). 
Increasing noise strength can enhance group-level synchronization while reducing the degree of intra-group synchronization and increasing its fluctuations. 

For coupled identical oscillators, while intra-group coupling helps intra-group synchronization, it can hinder group-level synchronization. Nevertheless, the sharing of the common noise across groups still robustly induces group-level synchronization over a wide range of $c_{in}$.

\subsection{\label{subsec:cpld_wi}Two groups of coupled nonidentical oscillators}
\begin{figure*}
\centering
\epsfig{figure=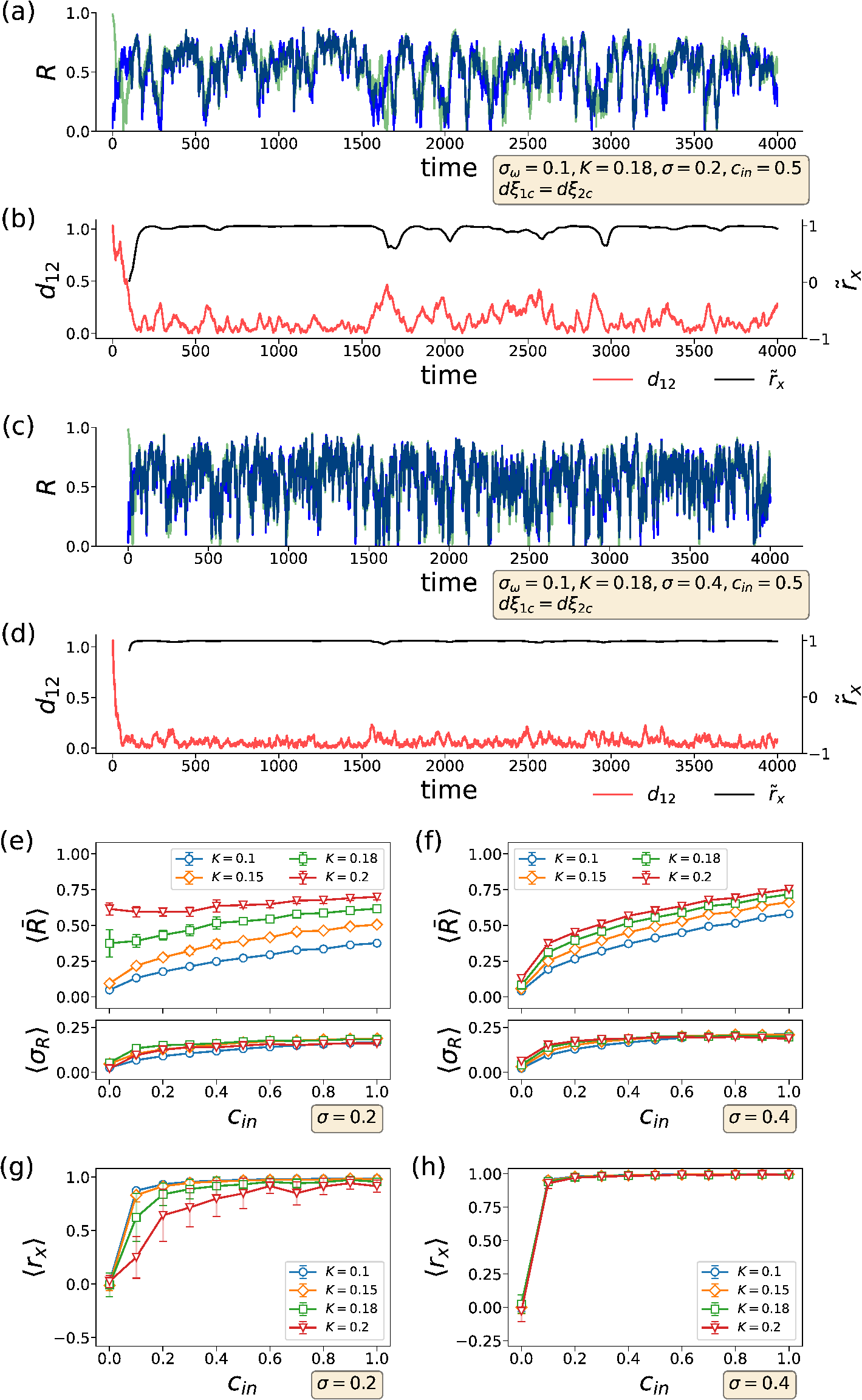, width= 13.8cm}
\caption{Two groups of coupled nonidentical oscillators ($\sigma_{\omega} \neq 0$, $K>0$) with the same common noise input to the groups ($d\xi_{1c} = d\xi_{2c}$): 
	(a) Synchronized behaviors of $R_1(t)$ and $R_2(t)$ with intermittent desynchrony for $\sigma_{\omega} = 0.1$, $K=0.18$, $\sigma = 0.2$, and $c_{in} = 0.5$.	
	(b) $d_{12}(t)$ and $\tilde r_x(t)$ for (a). 
	(c) Synchronized behaviors of $R_1(t)$ and $R_2(t)$ for $\sigma_{\omega} = 0.1$, $K=0.18$, $\sigma = 0.4$, and $c_{in} = 0.5$.	
	(d) $d_{12}(t)$ and $\tilde r_x(t)$ for (c). 
	$\langle \bar R \rangle$ and $\langle \sigma_R \rangle$ as functions of $c_{in}$ for (e) $\sigma=0.2$ 
	and (f) $\sigma=0.4$. 
	$\langle r_x \rangle$ as functions of $c_{in}$ for (g) $\sigma=0.2$ and (h) $\sigma=0.4$. 
	$\Delta(\theta) = -\sin\theta$ for all the figures.
	Initial conditions are as in Fig.~2.
	}
\label{fig:cpld_wi}
\end{figure*}

Finally, we study the case of two groups of coupled nonidentical oscillators, which combines frequency heterogeneity ($\sigma_\omega \neq 0$) with intra-group coupling ($K > 0$). As in the previous cases, we observe fluctuating order parameters of a single group and group-level synchronization between the two groups (Figs.~\ref{fig:cpld_wi}(a)-(d)). 
In this case, when we increase noise strength, the degree of group-level synchronization increases with diminished occurrence of brief desynchronization (Figs.~\ref{fig:cpld_wi}(b) and (d)).

Figures~\ref{fig:cpld_wi}(e) and (f) show how the degree of synchronization 
of a single group varies with $c_{in}$ and $K$ for two noise strengths.
Previous studies on coupled nonidentical oscillators showed that in the absence of noise input ($\sigma=0$), a single group of oscillators exhibits a phase transition toward synchronization as the coupling strength $K$ increases \cite{winfree2001,pikovsky_book,sync,kura,strogatz2000,kura_review}. 
The system shows incoherent behavior for $K<K_c$, and coherent behavior for $K>K_c$. 
Similarly, in this system with noise input ($\sigma>0$) and $c_{in} = 0$, there exists a critical $K_c$. At this critical point, the system transitions from an incoherent state where $R \approx 0$ to a synchronized state (Fig.~\ref{fig:cpld_wi}(e)).

	For fixed values of $\sigma$ and $K$, $\langle \bar R\rangle$ typically monotonically increases from (non)zero values to finite values less than $1$ as $c_{in}$ increases from $0$ to $1$ (Figs.~\ref{fig:cpld_wi}(e) and (f)) while the curves for $\langle \sigma_R \rangle$ are concave functions which first increase from near zero and then saturate at finite values.
	When we vary coupling strength $K$ at fixed $\sigma$ and $c_{in}$, $\langle \bar R\rangle$ increases while $\langle \sigma_R \rangle$ decreases.
	Increasing noise strength $\sigma$ can decrease or increase $\langle \bar R\rangle$ for fixed values of $K$ and $c_{in}$, depending on the specific parameter values.

As in the cases of coupled identical oscillators, group-level synchronization shows similar dependencies on parameters $K$ and $\sigma$ (Figs.~\ref{fig:cpld_wi}(g)-(h)).
With small noise strength $\sigma$, larger coupling strength enhances intra-group synchronization but hinders group-level synchronization (Fig.~\ref{fig:cpld_wi}(g)). Nevertheless, we observe group-level synchronization over a wide range of $c_{in}$ values for $K=0.1$, $0.15$, and $0.18$.
With larger noise strength $\sigma=0.4$ and $K$ values of Fig.~\ref{fig:cpld_wi}(h), strong group-level synchronization persists down to $c_{in} \approx 0.1$. 

For coupled nonidentical oscillators, while both frequency heterogeneity and intra-group coupling affect intra-group synchronization and its fluctuations, the sharing of the common noise across groups still induces group-level synchronization over a wide range of $c_{in}$. Strong intra-group coupling promoting intra-group synchronization can hinder group-level synchronization.

In this section, through numerical simulations, we examine four cases of the model:
uncoupled identical oscillators, uncoupled nonidentical oscillators, coupled
identical oscillators, and coupled nonidentical oscillators. We show that
significant fluctuations in the degree of intra-group synchronization occur and group-level synchronization between the two groups in the absence of inter-group coupling
is observed when groups share
common noise input in the presence of uncorrelated local noise.
The robustness of this group-level synchronization across different system sizes is demonstrated in Appendix C.

\begin{figure*}
\centering
\epsfig{figure=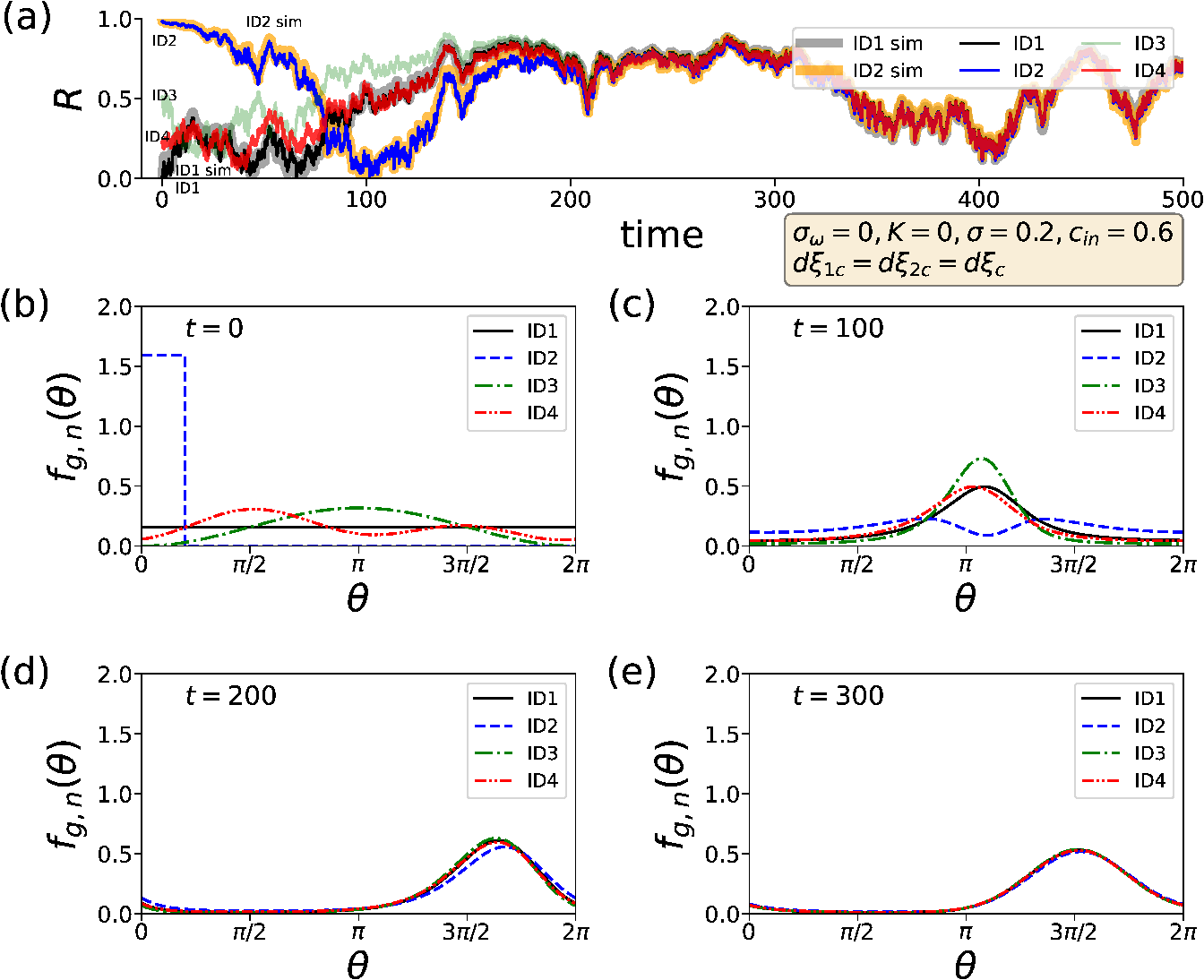, width= 13.8cm}
	\caption{$R(t)$ and $f_{g,n}(\theta)$ for two groups of uncoupled identical oscillators ($\sigma_\omega = 0$, $K = 0$) with the same common noise input to the groups ($d\xi_{1c} = d\xi_{2c}$): 
(a) $R(t)$ from simulations of the two groups with different initial conditions (ID1\:sim, ID2\:sim) and from numerical evolution of $f_{g,n}(\theta)$ through the mapping of Eq.~(\ref{eq:f_np1}) for different initial conditions (ID1, ID2, ID3, ID4).
$f_{g,n}(\theta)$ at (b) $t = 0$, (c) $t = 100$, (d) $t = 200$, and (e) $t = 300$.
$\sigma=0.2$, $c_{in} = 0.6$, and $\Delta(\theta) = -\sin\theta$.
Initial phase densities $f_{g,0}(\theta) = 1/(2\pi)$ for ID1;
$f_{g,0}(\theta) = \chi_{[0,0.2\pi)}(\theta)/(0.2\pi)$ for ID2, where $\chi_A(x) = 1$ for $x \in A$, otherwise $0$;
$f_{g,0}(\theta) = (-\cos\theta+1)/(2\pi)$ for ID3;
$f_{g,0}(\theta) = (-\cos(2\theta)+0.9\cos(\theta-0.6\pi)+2)/(4\pi)$ for ID4.
}
\label{fig:R_f_ucpld_w}
\end{figure*}

\begin{figure*}
\centering
\epsfig{figure=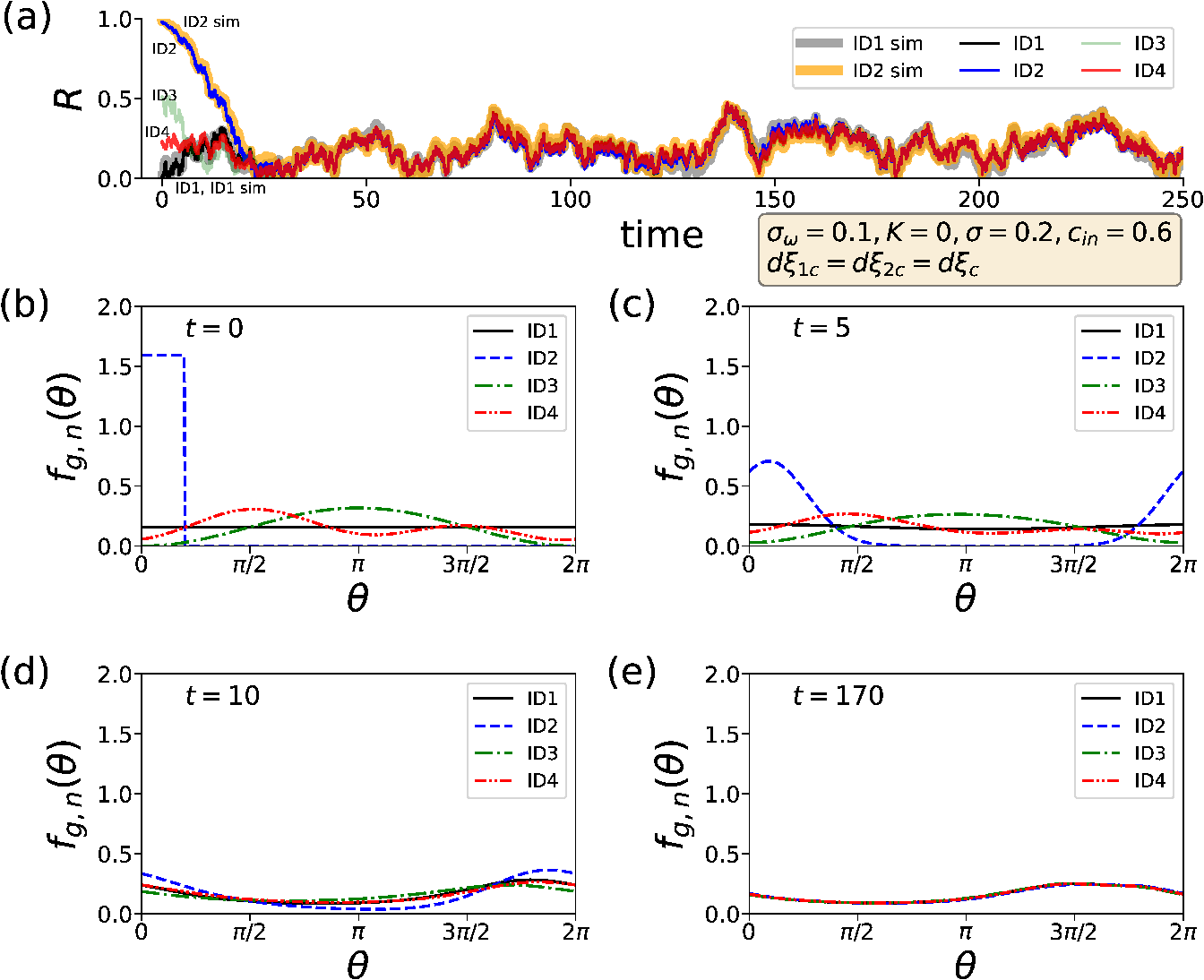, width= 13.8cm}
\caption{$R(t)$ and $f_{g,n}(\theta)$ for two groups of uncoupled nonidentical oscillators ($\sigma_\omega = 0.1$, $K = 0$) with the same common noise input to the groups ($d\xi_{1c} = d\xi_{2c}$):  
(a) $R(t)$ from simulations of the two groups with different initial conditions (ID1\:sim, ID2\:sim) and from numerical evolution of $f_{g,n}(\theta)$ for different initial conditions (ID1, ID2, ID3, ID4).
$f_{g,n}(\theta)$ at (b) $t=0$, (c) $t=5$, (d) $t=10$, and (e) $t=170$. $\sigma=0.2$, $c_{in}=0.6$, and  $\Delta(\theta) = -\sin\theta$.
	Other details are as in Fig.~\ref{fig:R_f_ucpld_w}.
}
  \label{fig:R_f_ucpld_wi}
\end{figure*}

\begin{figure*}
\centering
\epsfig{figure=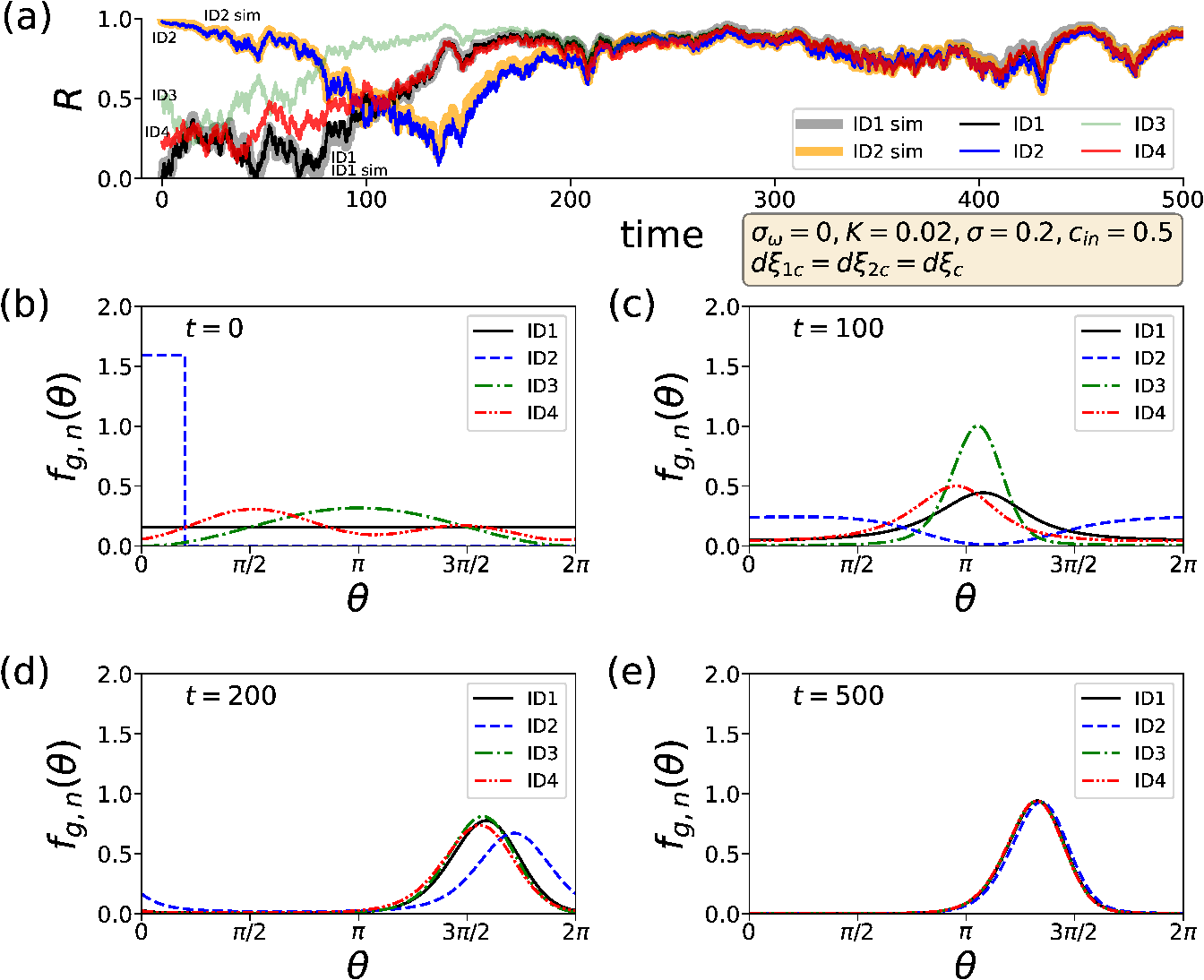, width= 13.8cm}
  \caption{$R(t)$ and $f_{g,n}(\theta)$ for two groups of coupled identical oscillators ($\sigma_\omega=0$, $K=0.02$) with the same common noise input to the groups ($d\xi_{1c} = d\xi_{2c}$): 
(a) $R(t)$ from simulations of the two groups with different initial conditions (ID1\:sim, ID2\:sim) and from numerical evolution of $f_{g,n}(\theta)$ for different initial conditions (ID1, ID2, ID3, ID4).
$f_{g,n}(\theta)$ at (b) $t=0$, (c) $t=100$, (d) $t=200$, and (e) $t=500$. $\sigma=0.2$, $c_{in}=0.5$, and $\Delta(\theta) = -\sin\theta$.
	Other details are as in Fig.~\ref{fig:R_f_ucpld_w}.
}
\label{fig:R_f_cpld_w}
\end{figure*}

\begin{figure*}
\centering
\epsfig{figure=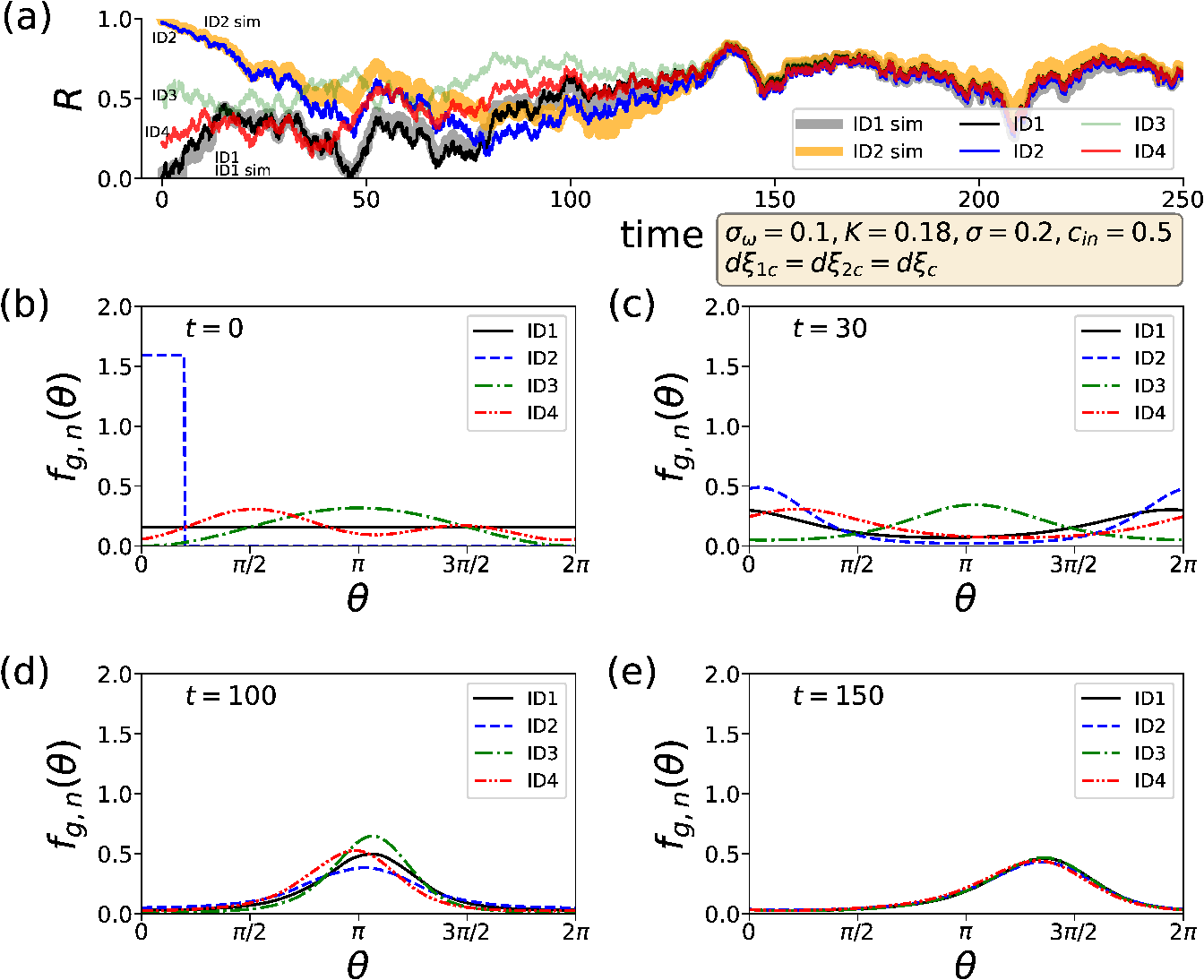, width= 13.8cm}
	\caption{$R(t)$ and $f_{g,n}(\theta)$ for two groups of coupled nonidentical oscillators ($\sigma_\omega=0.1$, $K=0.18$) with the same common noise input to the groups ($d\xi_{1c} = d\xi_{2c}$): 
(a) $R(t)$ from simulations of the two groups with different initial conditions (ID1\:sim, ID2\:sim) and from numerical evolution of $f_{g,n}(\theta)$ for different initial conditions (ID1, ID2, ID3, ID4).
$f_{g,n}(\theta)$ at (b) $t=0$, (c) $t=30$, (d) $t=100$, and (e) $t=150$.  
$\sigma=0.2$, $c_{in}=0.5$, and $\Delta(\theta) = -\sin\theta$.
	Other details are as in Fig.~\ref{fig:R_f_ucpld_w}.
}
\label{fig:R_f_cpld_wi}
\end{figure*}

\section{Analysis}
In this section, we derive a discrete-time mapping for the phase distribution evolution using the discretization of the model through the Euler-Maruyama method.
The group-level synchronization is shown again by numerically evolving the phase distributions of the groups. This matches well with the simulation results.
In the cases of uncoupled oscillators, we analytically explain how the group-level synchronization is achieved using the phase distribution evolution mapping.

Discretization of Eq.~(\ref{eq:model}) through the Euler-Maruyama method with time step $\delta t$ gives the following approximate mapping:
\begin{flalign}
	\quad\theta_{gi,n+1} &= \theta_{gi,n}+\biggl [\omega_{gi}
    +\frac{K \Bigl(Z_{g,n} {\rm e}^{-\mathrm{i}\theta_{gi,n}}-Z_{g,n}^* {\rm e}^{\mathrm{i}\theta_{gi,n}}\Bigr)}{2\mathrm{i}}  \nonumber \\
	&\quad+\frac{\sigma^2}{2}\Delta(\theta_{gi,n}) \Delta'(\theta_{gi,n})\biggr] \delta t &\nonumber \\ 
	&\quad+\sigma \Delta \left(\theta_{gi,n}\right)\bigl(\sqrt{c_{in}}\lambda_{gc,n}+\sqrt{1-c_{in}}\lambda_{gi,n}\bigr) \nonumber \\ 
& ~n = 0, 1, 2, ...,
\label{eq:theta_map}
\end{flalign}
where $\theta_{gi,n}$ stands for the phase of oscillator $i$ of group $g$ at time $t=n\delta t$. 
Here, we use $Z_{g,n}$ which is the complex order parameter defined in Eq.~(\ref{eq:op}) at time $t_n$ to represent the coupling term of Eq.~(\ref{eq:model}). 
The random variables $\lambda_{gc,n}$ and $\lambda_{gi,n}$ are independent and identically distributed normal random variables with expected value zero and variance $\delta t$.
For given values of $K$ and $\sigma$, $\theta_{gi,n+1}$ is given by a mapping $\theta_{gi,n+1} = \gamma(\theta_{gi,n};\omega_{gi},Z_{g,n},\lambda_{gc,n},\lambda_{gi,n})$. 

In the limit of $N\rightarrow \infty$, we can define a phase density function (phase density) $f_{g,n}(\theta)$ for a given set of parameters such that $f_{g,n}(\theta)d\theta$ denotes the fraction of oscillators of group $g$ whose phases lie in $[\theta, \theta + d\theta]$ at time $t=n\delta t$.

Considering that $f_{g,n+1}(\theta)$ is the phase density for $\theta_{gi,n+1}$  which evolves from $\theta_{gi,n}$ through the mapping $\gamma$, we can find equations describing the time evolution of the phase distribution as follows. 
\begin{subequations}
	\begin{flalign}
      \label{eq:f_np1_a}
      &f_{g\omega,n+1}(\theta) &\nonumber \\ 
      &= \int_{-\infty}^{+\infty} d\lambda  
        \frac{h(\lambda) f_{g\omega,n}\left(\gamma_{g\omega,n}^{-1}(\theta;Z_{g,n},\lambda_{gc,n},\lambda)\right)}{\gamma_{g\omega,n}'\left(\gamma_{g\omega,n}^{-1}(\theta;Z_{g,n},\lambda_{gc,n},\lambda);Z_{g,n},\lambda_{gc,n},\lambda\right)}, \\
      \label{eq:f_np1_b}
      &f_{g,n+1}(\theta) = \int_{0}^{+\infty} d\omega ~p(\omega) f_{g\omega,n+1}(\theta),
	\end{flalign}
\label{eq:f_np1}
\end{subequations}
where $f_{g\omega,n}(\theta)$ represents the phase density defined such that $f_{g\omega,n}(\theta)d\theta$ denotes the fraction of oscillators whose phases lie in $[\theta, \theta+d\theta]$ at time $t=n\delta t$ among the oscillators of group $g$ whose natural frequencies are in $[\omega,\omega+d\omega]$. 
The variable $\lambda$ represents the local noise input to individual oscillators, corresponding to $\lambda_{gi,n}$ in Eq.~(\ref{eq:theta_map}).
In Eq.~(\ref{eq:f_np1_a}), $\gamma_{g\omega,n}$ is the mapping from $\theta_{gi,n}$ to $\theta_{gi,n+1}$ given by
\begin{flalign}
  &\gamma_{g\omega,n}(\theta;Z_{g,n},\lambda_{gc,n},\lambda) \nonumber \\
    &~~= \theta + \biggl [\omega+\frac{K\Bigl(Z_{g,n} {\rm e}^{-\mathrm{i}\theta}-Z_{g,n}^* {\rm e}^{\mathrm{i}\theta}\Bigr)}{2\mathrm{i}} +\frac{\sigma^2}{2}\Delta(\theta) \Delta'(\theta) \biggr ]\delta t &\nonumber \\
    &~~\quad+\sigma\Delta(\theta)\bigl(\sqrt{c_{in}}\lambda_{gc,n}+\sqrt{1-c_{in}}\lambda\bigr), 
\end{flalign}
where the complex order parameter $Z_{g,n}$ is obtained from the phase density as $Z_{g,n} = \int_{0}^{2\pi}d\theta f_{g,n}(\theta) {\rm e}^{\mathrm{i}\theta}$, $\gamma_{g \omega,n}^{-1}$ is the inverse function of $\gamma_{g\omega,n}$, and $\gamma_{g\omega,n}' = d \gamma_{g\omega,n}/ d \theta$. 
For a given $\lambda$, the right side of Eq.~(\ref{eq:f_np1_a}) is an $h$-weighted integral of the phase density transformed from $t=n\delta t$ to $t=(n+1)\delta t$. The weight function $h$ is a Gaussian probability density function $h(\lambda)=\frac{1}{\sigma_\lambda \sqrt{2\pi}}{\rm e}^{-\frac{1}{2}\lambda^2/\sigma_\lambda^2}$ with $\sigma_\lambda=\sqrt{\delta t}$.

As shown in Eq.~(\ref{eq:f_np1_b}), $f_{g,n+1}(\theta)$ is  finally obtained through a $p$-weighted integral of  $f_{g\omega,n+1}(\theta)$, where  $p(\omega)$ is the probability density function for $\omega$.  
	For identical oscillators, $p(\omega)=\delta(\omega-\omega_0)$, a Dirac delta function, is used, while for nonidentical oscillators, $p(\omega)=\frac{1}{\sigma_\omega \sqrt{2\pi}}{\rm{e}}^{-\frac{1}{2} {(\omega-\omega_0)^2}/{\sigma_\omega^2}}$ is used.

Note that a group of nonidentical oscillators can be regarded as being composed of
subgroups of identical oscillators, with each subgroup having a different natural frequency.
The case of a group of identical oscillators is a special case with one subgroup.
Equation (\ref{eq:f_np1_a}) describes the evolution of the phase density for the subgroup with frequency $\omega$. Equation (\ref{eq:f_np1_b}) gives the group-level phase density $f_{g,n+1}(\theta)$ obtained by integrating the subgroup-level phase densities $f_{g\omega,n+1}(\theta)$ weighted by the probability density function $p(\omega)$ for the natural frequencies. 

We numerically compute the evolution of the phase distributions using Eqs.~(\ref{eq:f_np1_a}) and (\ref{eq:f_np1_b}), and the results from this evolution match well with those from the simulations of Eq.~(\ref{eq:model}).
We obtain $\theta^{old}=\gamma^{-1}_{g\omega,n}(\theta;Z_{g,n},\lambda_{gc,n},\lambda)$, the previous phase that evolves to the current phase $\theta$ under the mapping $\gamma_{g\omega,n}$, using a fixed-point iteration method:
\begin{eqnarray}
\theta^{old}_{m+1} = \theta + \theta^{old}_m
- \gamma_{g\omega,n}(\theta^{old}_m;Z_{g,n},\lambda_{gc,n},\lambda)
\end{eqnarray}
with $\theta_0=\theta$ as the initial guess. 
The integrations of Eqs.~(\ref{eq:f_np1_a}) and (\ref{eq:f_np1_b}) are calculated using composite Simpson's $1/3$ rule \cite{suli2003}.

In Fig.~\ref{fig:R_f_ucpld_w}(a), we show time series for $R$ of two groups of uncoupled identical oscillators with the same common noise input to the groups. 
The time series $R(t)$ of the two groups (ID1\:sim and ID2\:sim) are obtained from the simulation of the model Eq.~(\ref{eq:model}) and four time series $R(t)$ are obtained from the numerical evolutions of $f_{g,n}(\theta)$ using Eq.~(\ref{eq:f_np1}) for four different initial conditions (ID1, ID2, ID3, and ID4).  
All the time series for $R$ eventually converge to the same trajectory, as in Fig.~\ref{fig:ucpld_w_two}.
The $R$ time series obtained from simulation and numerical evolution match well when starting from the same initial condition: ID1\:sim vs ID1, ID2\:sim vs ID2.
In Figs.~\ref{fig:R_f_ucpld_w}(b)-(e), we also show the corresponding evolutions of the phase distributions. 
The group-level synchronization is manifested as the synchronous evolution of the phase distributions. 
Similar results are shown in Figs.~\ref{fig:R_f_ucpld_wi}, \ref{fig:R_f_cpld_w}, and \ref{fig:R_f_cpld_wi} for cases of uncoupled nonidentical, coupled identical, and coupled nonidentical oscillators, respectively.

Now, let us analytically examine how the group-level synchronization, which is equivalent to the synchronous evolution of the phase distributions, is achieved.

First, we confirm that group-level synchronization represents an invariant state of the system. Note that this does not mean the state is stable. 
If two groups of (non)identical oscillators receiving shared common noise ($\lambda_{1c,n}=\lambda_{2c,n}=\lambda_{c,n}$) start with each corresponding pair of subgroups from the two groups having the same phase distribution as shown below 
\begin{eqnarray}
    f_{1\omega,n}(\theta)=f_{2\omega,n}(\theta) ~\textrm{for all $\omega$},
	\end{eqnarray}
	then, $f_{1,n}(\theta)=f_{2,n}(\theta)$ and thus $\gamma_{1\omega,n}(\theta)=\gamma_{2\omega,n}(\theta)$ for all $\omega$.
	Therefore, $f_{1\omega,n}(\theta) = f_{2\omega,n}(\theta)$ for all $\omega$ and $f_{1,n}(\theta)=f_{2,n}(\theta)$ hold for all subsequent time steps.

To understand how initially different phase distributions converge to identical distributions, we can decompose each phase density $f_{g\omega,n}(\theta)$ into two parts: $f_{1\omega,n}(\theta) = f_{\omega,n}^c(\theta) + f_{1\omega,n}^d(\theta)$ and $f_{2\omega,n}(\theta) = f_{\omega,n}^c(\theta) + f_{2\omega,n}^d(\theta)$, where the overlap phase density $f_{\omega,n}^c(\theta)$ is defined as $f_{\omega,n}^c(\theta) \equiv \min\{f_{1\omega,n}(\theta),f_{2\omega,n}(\theta)\}$ and represents the overlapping portion between $f_{1\omega,n}(\theta)$ and $f_{2\omega,n}(\theta)$. 
The terms $f_{g\omega,n}^d(\theta)$ are the discrepancy phase densities,
representing the non-overlapping portions of each phase density.
Due to the normalization condition of the phase densities, one phase density cannot be lower than the other across the entire $\theta$ range. Consequently, there must be at least one crossing point between the two phase densities.

\begin{figure*}
\centering
\epsfig{figure=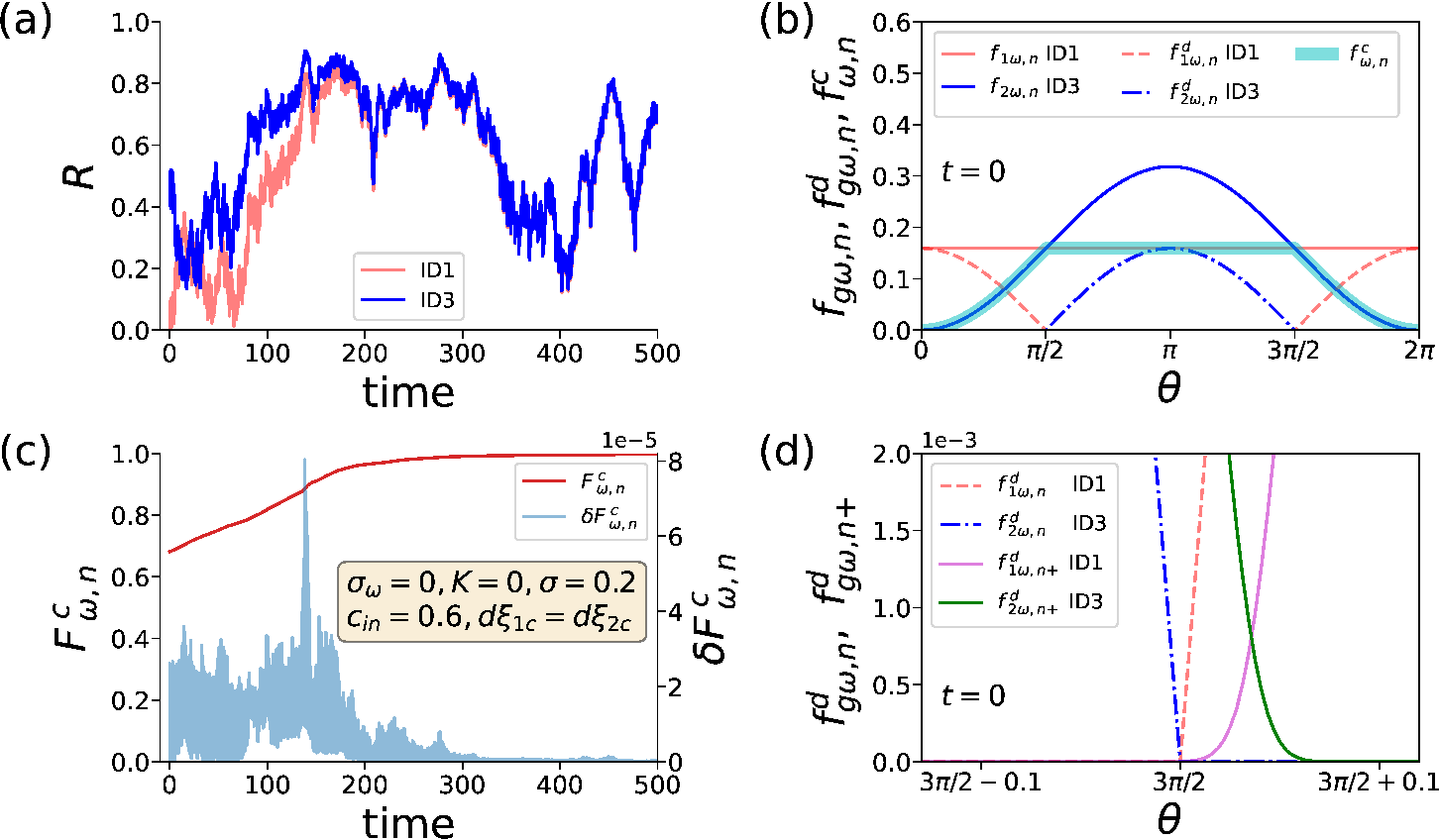, width= 13.8cm}
	\caption{Convergence of phase distributions for two groups of uncoupled identical oscillators ($\sigma_\omega=0$, $K=0$) with the same common noise input to the groups ($d\xi_{1c} = d\xi_{2c}$):   
(a) $R(t)$ from numerical evolution of $f_{g,n}(\theta)$ for the two groups starting with different initial conditions (ID1, ID2).
(b) Phase densities $f_{g\omega,n}(\theta)$, discrepancy phase densities $f_{g\omega, n}^d(\theta)$, overlap phase density $f_{\omega,n}^c(\theta)$ at $t=0~(n=0)$.
(c) Evolution of overlap fraction $F_{\omega,n}^c$ and its incremental change $\delta F_{\omega,n}^c$. 
(d) Overlap increase due to new overlap arising between evolved phase densities $f_{g\omega, n+}^d(\theta)$ from the non-overlapping discrepancy phase densities $f_{g\omega,n}^d(\theta)$ at $t=0$. 
$\sigma=0.2$, $c_{in}=0.6$, and $\Delta(\theta) = -\sin\theta$.
For the definitions of the terms, see the text.  Other details are as in Fig.~\ref{fig:R_f_ucpld_w}.
}
\label{fig:R_f_ucpld_w_F}
\end{figure*}

We now focus on the case of two groups of uncoupled identical oscillators.

The overlap phase density at subsequent time points contains the phase density that results from the evolution of the current overlap phase density $f_{\omega,n}^c(\theta)$. This occurs because the evolution from $f_{\omega,n}^c(\theta)$ gives the identical phase density for both groups, following the map described in Eq.~(\ref{eq:f_np1_a}) with $\gamma_{g\omega,n}$ identical for both groups.
Therefore, in this case the overlap fraction $F_{\omega,n}^c \equiv \int_0^{2\pi} d\theta f_{\omega,n}^c(\theta)$ does not decrease: $F_{\omega,n}^c \leq F_{\omega,n+1}^c$.

The discrepancy phase densities $f_{1\omega,n}^d(\theta)$ and $f_{2\omega,n}^d(\theta)$ intersect at a crossing phase $\theta^*$ where $f_{1\omega,n}(\theta^*) = f_{2\omega,n}(\theta^*)$, implying $f_{1\omega,n}^d(\theta^*) = f_{2\omega,n}^d(\theta^*) = 0$. In the vicinity of $\theta^*$, these discrepancy phase densities exhibit complementary behavior: one discrepancy phase density is strictly greater than zero for $\theta < \theta^*$ and zero for $\theta > \theta^*$, while the other one shows the opposite behavior, being zero for $\theta < \theta^*$ and strictly greater than zero for $\theta > \theta^*$.

To understand how the overlap increases, it is sufficient to consider the following case in the vicinity of $\theta^*$: $f_{1\omega,n}^d(\theta) > 0$ and $f_{2\omega,n}^d(\theta) = 0$ for $\theta < \theta^*$; $f_{1\omega,n}^d(\theta) = 0$ and $f_{2\omega,n}^d(\theta) > 0$ for $\theta > \theta^*$. 
For oscillators belonging to the fraction represented by $f_{g\omega,n}^d(\theta)$ near $\theta^*$, we examine the phase evolution of a representative oscillator $gl$. 
At time point $n$, its phase is $\theta^*+\epsilon_{gl}$ with $|\epsilon_{gl}|\ll 1$, $\epsilon_{1l}<0$  and $\epsilon_{2l}>0$. 
With $\lambda_{1c,n}=\lambda_{2c,n} = \lambda_{c,n}$ for all $n$, the phase at time point $n+1$ is given by:
\begin{subequations} 
\begin{align}
		\theta_{gl,n+1} &= 
        (\theta^*\!+\!\epsilon_{gl}) + \biggl [\omega + \frac{\sigma^2}{2}\Delta(\theta^*\!+\!\epsilon_{gl})\Delta'(\theta^*\!+\!\epsilon_{gl})\biggr ]\delta t &\nonumber \\
        &\quad+ \sigma\Delta(\theta^*+\epsilon_{gl})\Bigl(\sqrt{c_{in}}\lambda_{c,n}+\sqrt{1-c_{in}}\lambda_{gl,n}\Bigr) \nonumber \\
		&\approx \theta^* + \biggl [\omega + \frac{\sigma^2}{2}\Delta(\theta^*)\Delta'(\theta^*)\biggr ]\delta t \nonumber \\ 
		&\quad+ \sigma\Delta(\theta^*)\Bigl(\sqrt{c_{in}}\lambda_{c,n}+\sqrt{1-c_{in}}\lambda_{gl,n}\Bigr) \nonumber \\
        &\quad+ \epsilon_{gl} \biggl [1 + \frac{\sigma^2}{2}\Bigl ({\Delta'(\theta^*)}^2 + \Delta(\theta^*)\Delta''(\theta^*) \Bigr)\delta t \nonumber \\
		&\quad+ \sigma\Delta'(\theta^*)\Bigl(\sqrt{c_{in}}\lambda_{c,n}+\sqrt{1-c_{in}}\lambda_{gl,n}\Bigr)\biggr ].  
\end{align}
\end{subequations}

When the phases of some oscillators from group $1$ overtake those of some oscillators from group $2$ at time point $n+1$, overlapping occurs between the phase densities evolved from the discrepancy phase densities.  
To check this, we calculate the phase difference of the oscillators $p$ and $q$, belonging to groups $1$ and $2$, respectively, at time point $n+1$.
\begin{subequations} 
  \begin{align}
    &\quad\theta_{1p,n+1} - \theta_{2q,n+1} &\nonumber \\
    &\quad\quad\approx  \sigma\Delta(\theta^*)\sqrt{1-c_{in}}(\lambda_{1p,n}- \lambda_{2q,n})\nonumber \\
        &\quad\quad\quad+ \biggl [1 + \frac{\sigma^2}{2}\Bigl ({\Delta'(\theta^*)}^2 + \Delta(\theta^*)\Delta''(\theta^*) \Bigr)\delta t \nonumber \\
        &\quad\quad\quad\quad+ \sigma\Delta'(\theta^*)\sqrt{c_{in}}\lambda_{c,n}\biggr ] (\epsilon_{1p}- \epsilon_{2q}) \nonumber \\
        &\quad\quad\quad+ \sigma\Delta'(\theta^*)\sqrt{1-c_{in}}(\epsilon_{1p}\lambda_{1p,n}-\epsilon_{2p}\lambda_{2p,n}). 
  \end{align}
\label{eq:check_overtake}
\end{subequations}
When this phase difference is positive, an overtaking occurs.
The second term is always negative, since $[\cdots]$ is positive and $(\epsilon_{1p}- \epsilon_{2q})$ is negative. 
For the cases with $c_{in} < 1$, during the evolution of the phase densities, the first term can be positive and larger in magnitude than the second and third terms at $\theta^*$ with $\Delta(\theta^*)\neq 0$ for certain realizations of the local noise inputs, and thus can make the difference positive. 
Therefore, near the phase $\theta^*+\bigl [\omega + \frac{\sigma^2}{2}\Delta(\theta^*)\Delta'(\theta^*)\bigr]\delta t+\sigma\Delta(\theta^*)\sqrt{c_{in}}\lambda_{c,n}$, there can be overlap between $f_{1\omega,n+}^d(\theta)$ and $f_{2\omega,n+}^d(\theta)$ which evolve from $f_{1\omega,n}^d(\theta)$ and $f_{2\omega,n}^d(\theta)$, respectively, as follows:
\begin{subequations}
	\begin{flalign}
        \quad f_{1\omega,n+}^d(\theta) &\equiv \int_{-\infty}^{+\infty} d\lambda  
        \frac{h(\lambda) f_{1\omega,n}^d\!\left(\gamma_{1\omega,n}^{-1}(\theta; \lambda_{c,n}, \lambda)\right)}{\gamma_{1\omega,n}'\!\left(\gamma_{1\omega,n}^{-1}(\theta; \lambda_{c,n}, \lambda); \lambda_{c,n}, \lambda\right)}, &\\
        f_{2\omega,n+}^d(\theta) &\equiv \int_{-\infty}^{+\infty} d\lambda  
        \frac{h(\lambda) f_{2\omega,n}^d\!\left(\gamma_{2\omega,n}^{-1}(\theta; \lambda_{c,n}, \lambda)\right)}{\gamma_{2\omega,n}'\!\left(\gamma_{2\omega,n}^{-1}(\theta; \lambda_{c,n}, \lambda); \lambda_{c,n}, \lambda\right)}.
	\end{flalign}
\end{subequations}

The area of the overlap between $f_{1\omega,n+}^d(\theta)$ and $f_{2\omega,n+}^d(\theta)$ corresponds to the overlap increase $\delta F_{\omega,n} \equiv F_{\omega,n+1}^c - F_{\omega,n}^c$. Eventually, the two phase densities $f_{1\omega,n}(\theta)$ and $f_{2\omega,n}(\theta)$ become fully overlapping. 
For very low $c_{in}$ values, full overlap of the two phase densities, which are nearly uniform over the phase range $[0,2\pi]$, does not correspond to group-level synchronization, since both $R_1$ and $R_2$ come to have near zero values. 
Meanwhile, for larger values of $c_{in}$ but less than $1$, $R$ values fluctuate and group-level synchronization of the two groups is observed with the full overlap. 
For the case with $c_{in}=1$, only the second term, which is always non-positive, remains in Eq.~(\ref{eq:check_overtake}) and thus there cannot be overlap between $f_{1\omega,n+}^d(\theta)$ and $f_{2\omega,n+}^d(\theta)$. 
Consequently, there is no increase in the overlapping portion between $f_{1\omega,n}(\theta)$ and $f_{2\omega,n}(\theta)$ as time progresses.
However, group-level synchronization between the two groups occurs without increase in the overlapping, because both groups come close to a fully synchronized state within each group with their phase densities sharply peaked around nearly identical phases at each time point as all the oscillators within each group receive the same noise input and this common noise is shared between the two groups. 
This synchronization of uncoupled identical oscillators within a group under common noise input has been demonstrated in Refs.~\cite{teramae2004,galan2006,galan2007a,galan2007b,marella2008,abouzeid2009}.

To confirm our analysis of how the overlap of the phase distributions of two groups of uncoupled identical oscillators occurs when the groups receive the same common noise input, we examine the phase distribution evolution of the two groups starting from two different initial conditions using the mapping of Eq.~(\ref{eq:f_np1}) in Fig.~\ref{fig:R_f_ucpld_w_F}. 
Figure~\ref{fig:R_f_ucpld_w_F}(a) displays $R(t)$ of the two groups obtained from the evolving phase distributions at time $t$ for $c_{in}=0.6$. 
The phase densities $f_{1\omega,n}(\theta)$ and $f_{2\omega,n}(\theta)$ shown in Fig.~\ref{fig:R_f_ucpld_w_F}(b) are the initial densities used for the numerical evolution. 
Note that in the cases of identical oscillators, it is sufficient to consider $f_{g\omega,n}(\theta)$ only since $f_{g,n}(\theta) = f_{g\omega,n}(\theta)$.
As explained above, the phase densities can be decomposed into two parts: the discrepancy phase densities $f_{g\omega,n}^d(\theta)$ (red and blue broken lines) and the overlap phase density $f_{\omega,n}^c(\theta)$ (cyan thick line).
In Fig.~\ref{fig:R_f_ucpld_w_F}(c), the overlap fraction $F_{\omega,n}^c$ monotonically increases with $\delta F_{\omega,n}^c \geq 0$ and saturates to $1$ as the two time series of $R$ come to fluctuate together. 
Figure~\ref{fig:R_f_ucpld_w_F}(d) shows how the overlap increases: new overlap emerges  between phase densities (purple and green lines) evolving from the non-overlapping discrepancy phase densities (red and blue broken lines).
These results confirm that the overlap mechanism accurately describes how group-level synchronization develops between two groups of uncoupled identical oscillators.

For the case of two groups of uncoupled nonidentical oscillators, the group-level synchronization can be explained using the results for groups of uncoupled identical oscillators. 
The frequency range of oscillators in each group can be subdivided into narrow,
non-overlapping sub-ranges. Oscillators within each sub-range form a subgroup
of almost identical oscillators.
When two such groups start from different initial conditions, each pair of corresponding subgroups from the two groups, consisting of oscillators in the same frequency sub-range, eventually achieves group-level synchronization as previously explained. Thus, through the synchronization of all such pairs of subgroups, the two groups achieve group-level synchronization.

Our analysis shows that groups of uncoupled oscillators without inter-group coupling achieve group-level synchronization under the same common noise in the presence of local noise inputs in the limit $N \rightarrow \infty$.

Analyzing phase distribution evolution with groups of coupled oscillators presents unique challenges beyond those encountered with groups of uncoupled oscillators.
In groups of coupled oscillators, the mapping $\gamma_{g\omega,n}$ depends on the group-specific complex order parameter $Z_{g,n}$ which can differ between groups and change as $n$ increases. 
This makes it difficult to examine the overlap of the phase distributions in contrast to the analysis with uncoupled oscillator cases.
Analytically explaining this group-level synchronization in the presence of intra-group coupling remains an open challenge for future research.
The Ott-Antonsen theory for describing complex Kuramoto order parameters \cite{lai2013,ott2008,pimenova2016,goldobin2019}  and stochastic Fokker-Planck equations for probability density evolution \cite{kawamura2008,bressloff2016}, both retaining common noise  terms, might provide foundations for deeper analytical understanding of group-level synchronization.

\section{Summary and Conclusions}
In summary, we have investigated noise-induced group-level synchronization between oscillator groups sharing the same common noise in the absence of inter-group coupling. Each group receives the same common noise as well as independent local noise inputs to individual oscillators. 
We have numerically studied both identical and nonidentical oscillators, with and without intra-group coupling.
Even though the four cases have different characteristics in terms of intra-group dynamics, we have found that the degree of synchronization within a single group typically exhibits significant temporal fluctuations. Importantly, our study has shown that when statistically equivalent groups share the same common noise, their collective oscillations become synchronized across the groups in the absence of inter-group coupling, regardless of their initial states.

This group-level synchronization between groups receiving the same common noise can be viewed as reliable reproduction of collective oscillations of a group in response to repeated application of a fluctuating input, regardless of its initial state and local noise inputs to the oscillators within the group. This interpretation is analogous to the correspondence between noise-induced synchronization of uncoupled oscillators and individual-level reliability of a single oscillator \cite{galan2007a}.

Using a phase density evolution mapping, we have analytically demonstrated how this group-level synchronization emerges between groups in the absence of intra-group coupling. The analysis has revealed that the overlap between phase distributions of different groups increases over time until the groups synchronize.

These findings extend our understanding of noise-induced synchronization from individual to group-level dynamics, revealing how shared noise can coordinate the collective behavior of otherwise independent oscillator groups.

This study may help understand correlated behaviors of neuronal groups or brain areas in the absence of direct coupling \cite{pessoa2014}. It can also provide insights into inter-subject brain wave synchronization observed when people watch the same movie \cite{hasson2004,hasson2009,denworth2023}. In these settings, the shared visual inputs may serve as a common fluctuating input to the brain of each observer.

Future work could extend this research by considering more complex connection networks among oscillators within each group while maintaining the absence of inter-group coupling. Additionally, examining cases where the common noise input affects only a subset of oscillators within each group could provide further insights.

\section*{Acknowledgements}
This work was supported by the National Institute for Mathematical Sciences (NIMS) grant funded by the Korean government (No. NIMS-B25910000). 

\renewcommand{\theequation}{A.\arabic{equation}}
\setcounter{equation}{0}  

\section*{Appendix A: Pearson correlation coefficients and Sliding window Pearson correlation coefficients}
The Pearson correlation coefficients used in Section \ref{sim_results} are calculated as follows.  For the simulation time range $[T_1, T_2]$, the Pearson correlation coefficients $r_x$ between the $x$-components $Z_{1x}$ and $Z_{2x}$ of complex order parameters 
and $r_y$ between the $y$-components $Z_{1y}$ and $Z_{2y}$ are given by
\begin{align}
	~~ r_x &\equiv \frac{\frac{1}{M}\sum_{i=1}^M \bigl [ Z_{1x}(t_i)-\overline{Z_{1x}}\,\bigr]\bigl[Z_{2x}(t_i)-\overline{Z_{2x}}\,\bigr]}{\sigma_{_{Z_{1x}}}\sigma_{_{Z_{2x}}}}, &\\
	r_y &\equiv \frac{\frac{1}{M}\sum_{i=1}^M \bigl [ Z_{1y}(t_i)-\overline{Z_{1y}}\,\bigr]\bigl[Z_{2y}(t_i)-\overline{Z_{2y}}\,\bigr]}{\sigma_{_{Z_{1y}}}\sigma_{_{Z_{2y}}}},
\end{align}
where $t_1 = T_1$ and $t_M = T_2$.
Here, $Z_{gx} = R_g \cos\Theta_g$ and $Z_{gy} = R_g \sin\Theta_g$.  $\overline{X}$ and $\sigma_X$ denote the time average and the standard deviation of $X(t)$, respectively. 

\renewcommand{\thefigure}{B.\arabic{figure}}
\setcounter{figure}{0}  
\begin{figure*}
\centering
\epsfig{figure=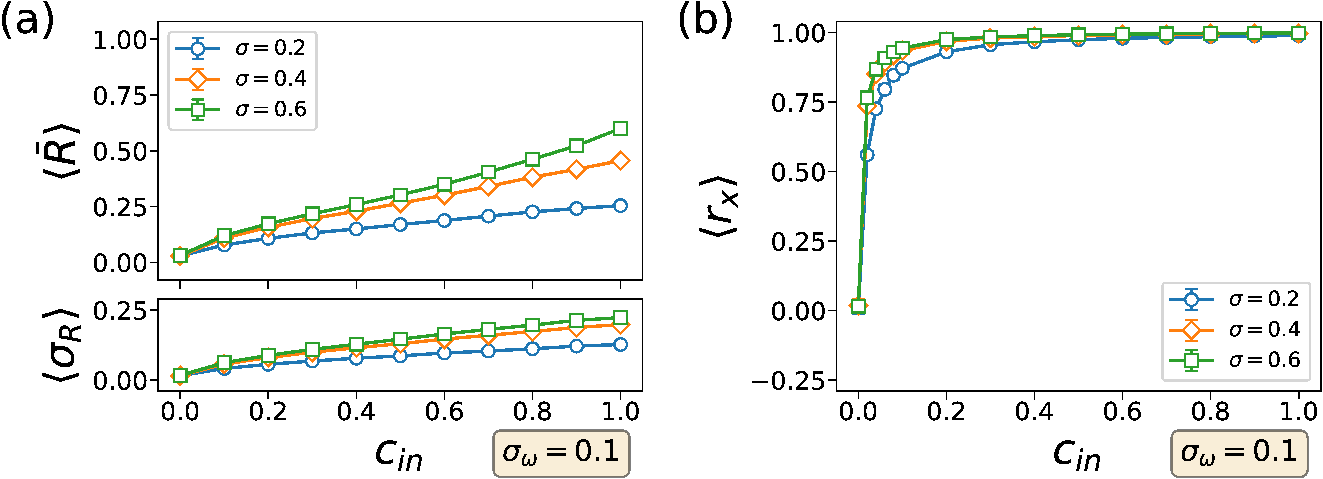, width= 13.8cm}
	\caption{Two groups of uncoupled nonidentical oscillators ($\sigma_{\omega} \neq 0$, $K=0$) with the same common noise input to the groups ($d\xi_{1c} = d\xi_{2c}$) : 
	(a) $\langle \bar R \rangle$, $\langle \sigma_R \rangle$, and (b) $\langle r_x \rangle$ as functions of $c_{in}$. $\sigma_{\omega} = 0.1$ and  $\Delta(\theta) = 1-\cos(\theta)$.
	Initial conditions are as in Fig.~2.
}
\label{fig:ucpld_wi_type1}
\end{figure*}

Since $R_g$ changes very slowly compared to the phase $\Theta_g$, for the time range $[T_1, T_2]$ with duration significantly longer than the oscillation period of the order parameters $Z_g=R_g {\rm e}^{\mathrm{i}\Theta_g}$,  $\overline {Z_{gx}} = \frac{1}{M}\sum_{i=1}^M R_g \cos(\Theta_g(t_i)) \approx 0$ and $\overline {Z_{gy}} =\frac{1}{M}\sum_{i=1}^M R_g \sin(\Theta_g(t_i)) \approx 0$. Similarly, the variances can be approximated as: 
\begin{align}
    {\sigma_{_{Z_{gx}}}}^2 &= \overline {{Z_{gx}}^2}-{\overline {Z_{gx}}}^2 \nonumber\\  
    &\approx  \frac{1}{M}\sum_{i=1}^M {R_g}^2(t_i) \cos^2(\Theta_g(t_i)) \nonumber \\ 
    &= \frac{1}{M}\sum_{i=1}^M {R_g}^2(t_i) \frac{1+ \cos(2\Theta_g(t_i))}{2} \nonumber \\  
    &\approx \frac{1}{M}\sum_{i=1}^M \frac{{R_g}^2(t_i)}{2},  \\ 
    {\sigma_{_{Z_{gy}}}}^2 &\approx 
    \frac{1}{M}\sum_{i=1}^M \frac{{R_g}^2(t_i)}{2}.   
\end{align}

Then, we have
\begin{align}
    r_x &\approx \frac{\frac{1}{M}\sum_{i=1}^M R_1(t_i)\cos\Theta_1(t_i)R_2(t_i)\cos\Theta_2(t_i)}{\sigma_{_{Z_{1x}}}\sigma_{_{Z_{2x}}}},~~~  \\
    r_y &\approx \frac{\frac{1}{M}\sum_{i=1}^M R_1(t_i)\sin\Theta_1(t_i)R_2(t_i)\sin\Theta_2(t_i)}{\sigma_{_{Z_{1y}}}\sigma_{_{Z_{2y}}}} \nonumber \\
    &=\frac{1}{M}\sum_{i=1}^M R_1(t_i)R_2(t_i)\Bigl [ \cos\Theta_1(t_i)\cos\Theta_2(t_i) \nonumber \\ 
    &\quad -\cos\bigl (\Theta_1(t_i)+\Theta_2(t_i)\bigl) \Bigl ]\times\frac{1}{\sigma_{_{Z_{1y}}}\sigma_{_{Z_{2y}}}} \nonumber \\
    &\approx r_x,
\end{align}
where we use $\frac{1}{M}\sum_{i=1}^M R_1(t_i)R_2(t_i)\cos\left (\Theta_1(t_i)+\Theta_2(t_i)\right) \approx 0$ due to slow changes in $R_g$ compared to the phases $\Theta_g$. 
This shows that $r_y \approx r_x$ for sufficiently long observation times.

Similarly, the sliding window Pearson correlation coefficients $\tilde {r}_x(t)$ between ${Z}_{1x}$ and ${Z}_{2x}$ 
and $\tilde {r}_y(t)$ between ${Z}_{1y}$ and ${Z}_{2y}$ 
over sliding time windows $[t-w,t]$ are defined. Through similar arguments as above, we find  
$\tilde {r}_y(t) \approx \tilde {r}_x(t)$.

\section*{Appendix B: Simulation Results of the model with type I phase resetting curve}
\begin{figure*}
\centering
\epsfig{figure=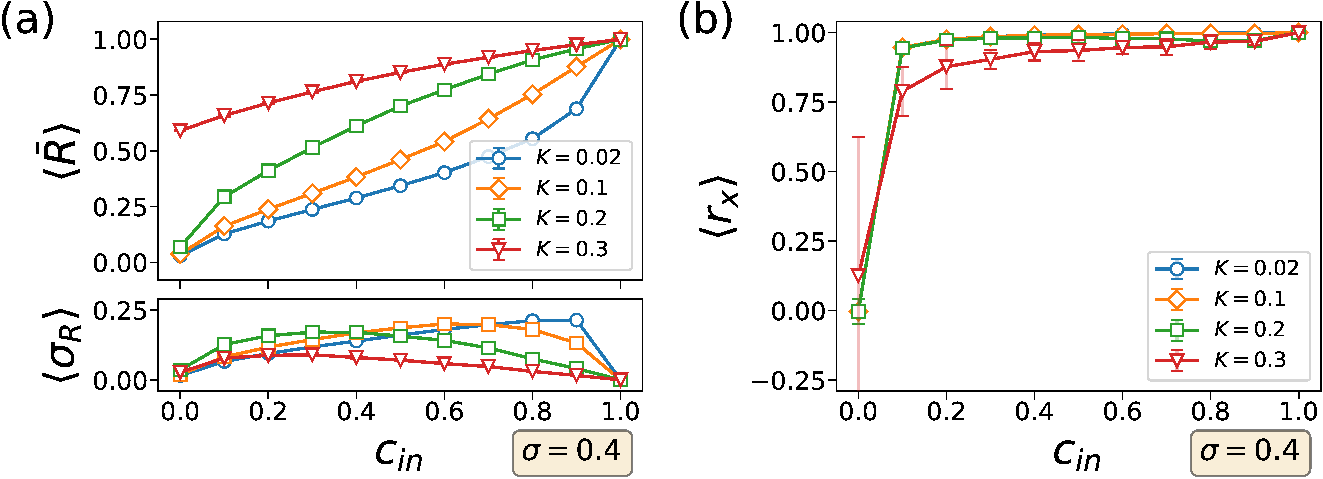, width= 13.8cm}
\caption{Two groups of coupled identical oscillators ($\sigma_{\omega} = 0$, $K>0$) with the same common noise input to the groups ($d\xi_{1c} = d\xi_{2c}$) : 
	(a) $\langle \bar R \rangle$, $\langle \sigma_R \rangle$, and (b) $\langle r_x \rangle$ as functions of $c_{in}$. $\sigma=0.4$ and $\Delta(\theta) = 1-\cos(\theta)$. 
	Initial conditions are as in Fig.~2.
    }
\label{fig:cpld_w_type1}
\end{figure*}

\begin{figure*}
\centering
\epsfig{figure=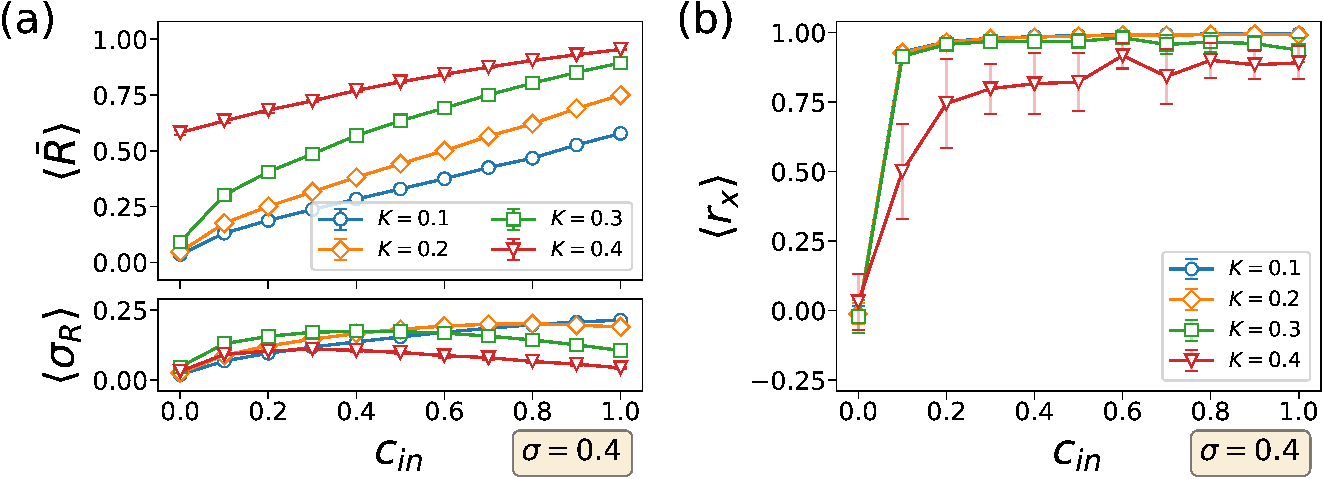, width= 13.8cm}
\caption{Two groups of coupled nonidentical oscillators ($\sigma_{\omega} \neq 0$, $K>0$) with the same common noise input to the groups ($d\xi_{1c} = d\xi_{2c}$) : 
	(a) $\langle \bar R \rangle$, $\langle \sigma_R \rangle$, and (b) $\langle r_x \rangle$ as functions of $c_{in}$. $\sigma=0.4$, $\sigma_\omega=0.1$, and $\Delta(\theta) = 1-\cos(\theta)$. 
	Initial conditions are as in Fig.~2.
    }
\label{fig:cpld_wi_type1}
\end{figure*}

In addition to the simulation results with type I PRC $\Delta(\theta)=1-\cos\theta$ for two groups of uncoupled identical oscillators shown in Sec.~\ref{sbsec:ucpld_w} (Figs.~\ref{fig:ucpld_w_one}(d) and \ref{fig:ucpld_w_two}(f)), here we present simulation results with the same PRC for the remaining cases involving two groups: uncoupled nonidentical oscillators (Fig.~\ref{fig:ucpld_wi_type1}), coupled identical oscillators (Fig.~\ref{fig:cpld_w_type1}), and coupled nonidentical oscillators (Fig.~\ref{fig:cpld_wi_type1}).
The simulation results are qualitatively similar to those with type II PRC in Sec.~\ref{sim_results} of the main text.

\section*{Appendix C: Simulations with different group sizes $N$}
To show that the group-level synchronization between oscillator groups in the absence of inter-group coupling is robustly observed, we perform simulations with different group sizes $N$ for a fixed noise strength $\sigma=0.2$. We consider three cases: two groups of uncoupled identical oscillators ($K=0$, Fig. \ref{fig:ucpld_w_N}), two groups of coupled identical oscillators ($K=0.02$, Fig. \ref{fig:cpld_w_N_K0p02}), and two groups of coupled identical oscillators ($K=0.1$, Fig. \ref{fig:cpld_w_N_K0p1}).
Each figure shows (a) synchronized time evolution of order parameters $R_1(t)$ and $R_2(t)$ for $N=1000$ and $N=10000$, (b) corresponding synchronization measure $d_{12}(t)$, (c) time-averaged measure $\overline{d_{12}}$ as a function of $c_{in}$ for different group sizes, and (d) scaling of $\overline{d_{12}} $ with group size $N$ demonstrating $1/\sqrt{N}$ behavior.
The consistent $1/\sqrt{N}$ behavior across  all coupling strengths studied confirms that the group-level synchronization observed in our study represents robust behavior that persists with large $N$.

\renewcommand{\thefigure}{C.\arabic{figure}}
\setcounter{figure}{0}  
\begin{figure*}
\centering
\epsfig{figure=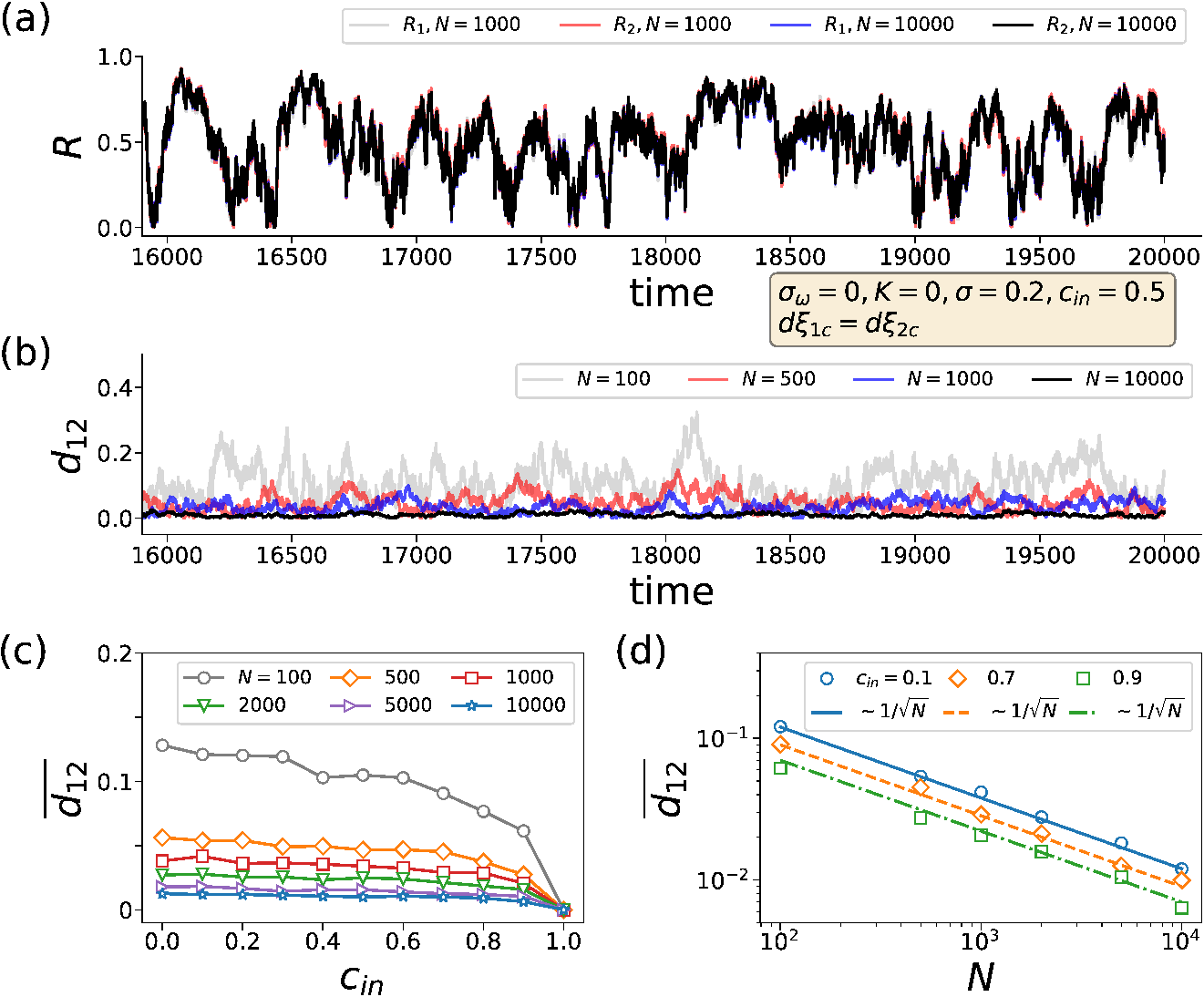, width= 13.8cm}
    \caption{Simulations with different group sizes $N$ for two groups of uncoupled identical oscillators ($\sigma_{\omega} = 0$, $K=0$) receiving the same common noise input ($d\xi_{1c} = d\xi_{2c}$): (a) Synchronized behaviors of $R_1(t)$ and $R_2(t)$ for $N=1000$ and $N=10000$ with $c_{in}=0.5$, using identical common noise realizations for the simulations. 
    (b) $d_{12}(t)$ defined in Eq.~(\ref{d12}) for (a) for different $N$ values. 
    (c) Time-averaged $\overline{d_{12}}$ as a function of $c_{in}$ for different $N$ values.  
    (d) $\overline{d_{12}}$ as a function of $N$ for fixed values of $c_{in}$, showing $1/\sqrt{N}$ scaling.
    Time averaging performed over $[T_1=4000,T_2=20000]$.
	$\sigma=0.2$ and $\Delta(\theta) = -\sin\theta$.
	Initial conditions are as in Fig.~2.
	}
\label{fig:ucpld_w_N}
\end{figure*}

\begin{figure*}
\centering
\epsfig{figure=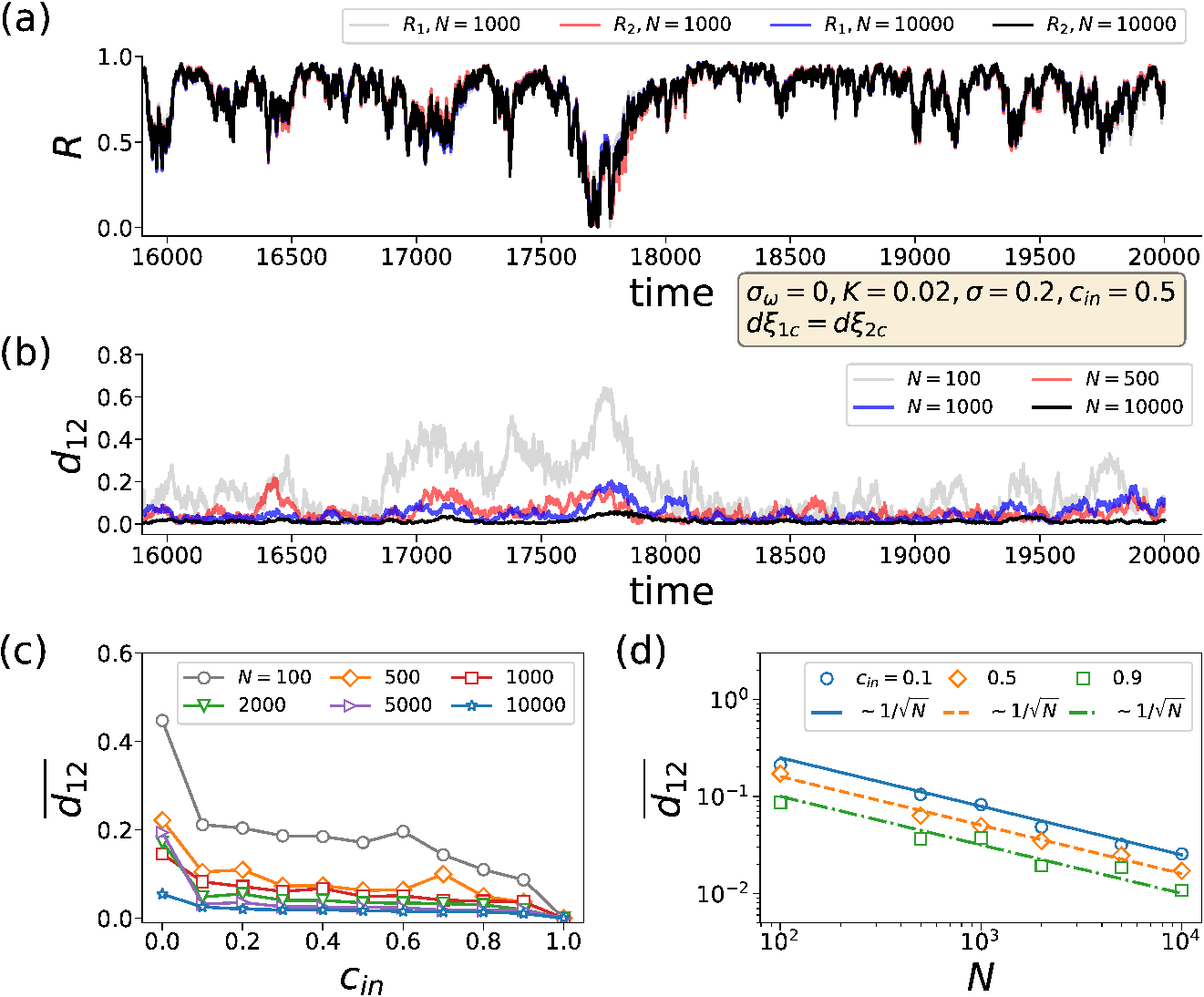, width= 13.8cm}
    \caption{Simulations with different group sizes $N$ for two groups of coupled identical oscillators ($\sigma_{\omega} = 0$, $K=0.02$) receiving the same common noise input ($d\xi_{1c} = d\xi_{2c}$): (a) Synchronized behaviors of $R_1(t)$ and $R_2(t)$ for $N=1000$ and $N=10000$ with $c_{in}=0.5$, using identical common noise realizations for the simulations. 
    (b) $d_{12}(t)$ defined in Eq.~(\ref{d12}) for (a) for different $N$ values. 
    (c) Time-averaged $\overline{d_{12}}$ as a function of $c_{in}$ for different $N$ values.  
    (d) $\overline{d_{12}}$ as a function of $N$ for fixed values of $c_{in}$, showing $1/\sqrt{N}$ scaling.
    Time averaging performed over $[T_1=4000,T_2=20000]$.
	$\sigma=0.2$ and $\Delta(\theta) = -\sin\theta$.
	Initial conditions are as in Fig.~2.
	}
\label{fig:cpld_w_N_K0p02}
\end{figure*}

\begin{figure*}
\centering
\epsfig{figure=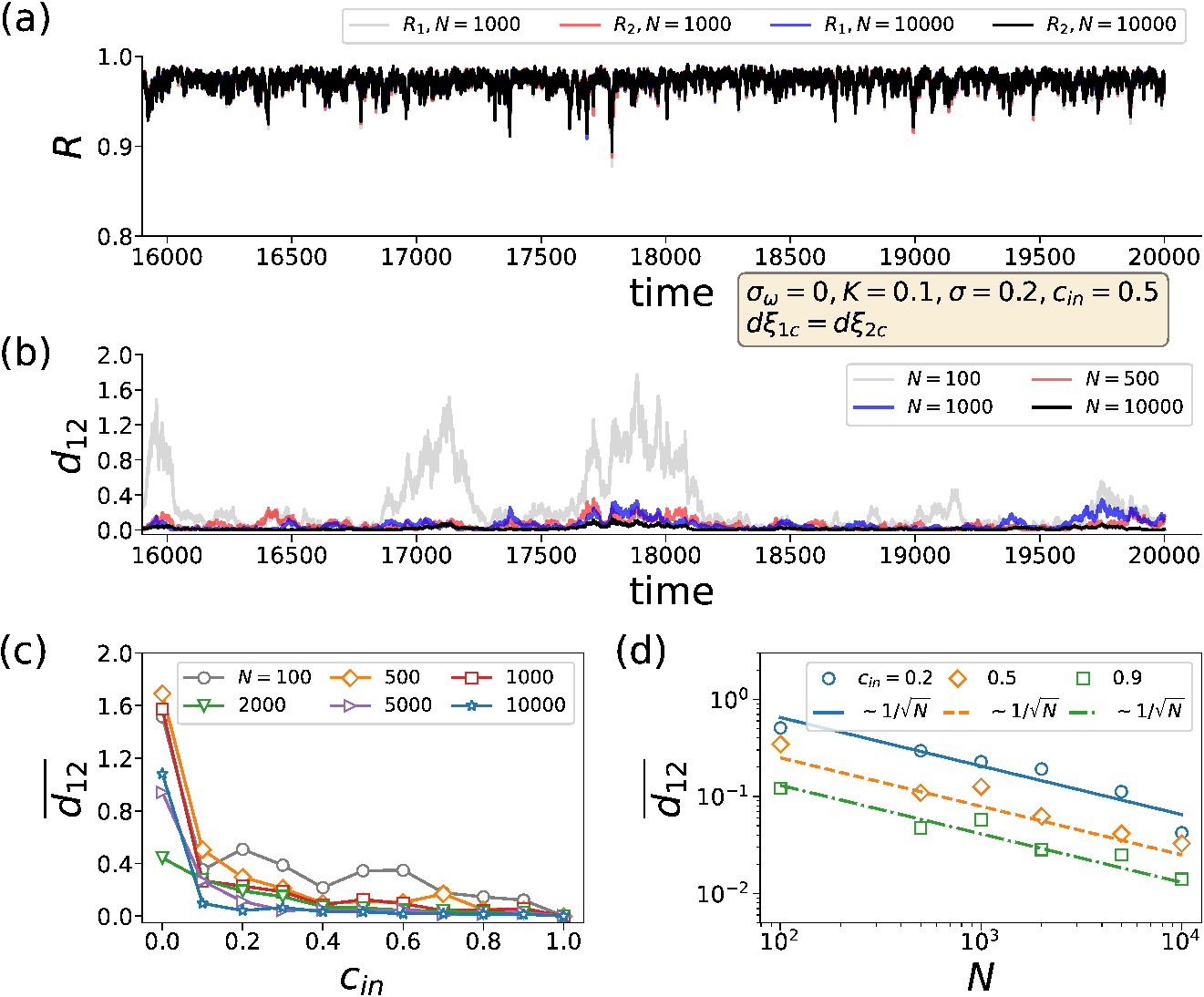, width= 13.8cm}
    \caption{Simulations with different group sizes $N$ for two groups of coupled identical oscillators ($\sigma_{\omega} = 0$, $K=0.1$) receiving the same common noise input ($d\xi_{1c} = d\xi_{2c}$): (a) Synchronized behaviors of $R_1(t)$ and $R_2(t)$ for $N=1000$ and $N=10000$ with $c_{in}=0.5$, using identical common noise realizations for the simulations. 
    (b) $d_{12}(t)$ defined in Eq.~(\ref{d12}) for (a) for different $N$ values. 
    (c) Time-averaged $\overline{d_{12}}$ as a function of $c_{in}$ for different $N$ values.  
    (d) $\overline{d_{12}}$ as a function of $N$ for fixed values of $c_{in}$, showing $1/\sqrt{N}$ scaling.
    Time averaging performed over $[T_1=4000,T_2=20000]$.
	$\sigma=0.2$ and $\Delta(\theta) = -\sin\theta$.
	Initial conditions are as in Fig.~2.
	}
\label{fig:cpld_w_N_K0p1}
\end{figure*}


\begin{thebibliography}{99}
\bibitem{winfree2001} A. T. Winfree, {\it The Geometry of Biological Time}, 2nd ed. (Springer-Verlag, New York, 2001). 
\bibitem{pikovsky_book} A. Pikovsky, M. Rosenblum, and J. Kurths, {\it Synchronization: A universal concept in nonlinear sciences} (Cambridge University Press, Cambridge, 2001)

\bibitem{sync} S. H. Strogatz, {\it Sync: The emergence science of spontaneous order} (Hyperion publisher, New York, 2003).

\bibitem{kura} Y. Kuramoto, {\it Chemical Oscillations, Waves, and Turbulence} (Springer, Berlin, 1984).

\bibitem{strogatz2000} S. H. Strogatz, From Kuramoto to Crawford: Exploring the onset of synchronization in populations of coupled oscillators, Physica D {\bf 143}, 1 (2000).

\bibitem{kura_review} J. A. Acebr{\'o}n {\it et al.}, 
The Kuramoto model: A simple paradigm for synchronization phenomena,
Rev. Mod. Phys. {\bf 77}, 137 (2005).

\bibitem{ermentrout2001} G. B. Ermentrout and D. Kleinfeld, Traveling Electrical Waves in Cortex : Insights from Phase Dynamics and Speculation on a Computational Role, Neuron {\bf 29}, 33 (2001).


\bibitem{pikovskii1984} A.S. Pikovskii, Synchronization and stochastization of array of self-excited oscillators by external noise, Radiophys.  Quantum Electron. {\bf 27}, 390 (1984).

\bibitem{goldobin2004} D.S. Goldobin and A. S. Pikovsky, Synchronization of periodic self-oscillations by common noise, Radiophys. Quantum Electron. {\bf 47}, 910 (2004).  

\bibitem{teramae2004} J. N. Teramae and D. Tanaka, 
Robustness of the Noise-Induced Phase Synchronization in a General Class of Limit Cycle Oscillators, Phys. Rev. Lett. {\bf 93}, 204103 (2004).

\bibitem{galan2006} R. F. Gal\'an, N. Fourcaud-Trocm\'e, G. B. Ermentrout, and N. N. Urban, Correlation-Induced Synchronization of Oscillations in Olfactory Bulb Neurons, J. Neurosci. {\bf 26}, 3646 (2006). 

\bibitem{galan2007a} R. F. Gal\'an, G. B. Ermentrout, and N. N. Urban,
Reliability and stochastic synchronization in type I vs. type II neural oscillators, Neurocomputing {\bf 70}, 2102 (2007). 

\bibitem{galan2007b} R. F. Gal\'an, G. B. Ermentrout, and N. N. Urban, 
	Stochastic dynamics of uncoupled neural oscillators: Fokker-Planck studies with the finite element method, Phys. Rev. E {\bf 76}, 056110 (2007).  

\bibitem{marella2008} S. Marella and G. B. Ermentrout, 
	Class-II neurons display a higher degree of stochastic synchronization than class-I neurons,
Phys. Rev. E {\bf 77}, 041918 (2008). 

\bibitem{abouzeid2009} A. Abouzeid and B. Ermentrout, 
Type-II phase resetting curve is optimal for stochastic synchrony, 
Phys. Rev. E {\bf 80}, 011911 (2009). 

\bibitem{goldobin2005} D. S. Goldobin and A. Pikovsky, Synchronization and desynchronization of self-sustained oscillators by common noise, Phys. Rev. E {\bf 71}, 045201(R) (2005).

\bibitem{goldobin2006} D.S. Goldobin and A. Pikovsky, Antireliability of noise-driven neurons, Phys. Rev. E {\bf 73}, 061906 (2006).

\bibitem{mainen1995} Z. Mainen and T. Sejnowski,  
Reliability of spike timing in neocortical neurons,
Science {\bf 268}, 1503 (1995).

\bibitem{nakao2007} H. Nakao, K. Arai, and Y. Kawamura, 
Noise-induced synchronization and clustering in ensembles of uncoupled limit-cycle oscillators,
Phys. Rev. Lett. {\bf 98}, 184101 (2007).

\bibitem{abouzeid2011} A. Abouzeid and B. Ermentrout, 
	Correlation transfer in stochastically driven neural oscillators over long and short time scales,
Phys. Rev. E {\bf 84}, 061914 (2011). 

\bibitem{nagai2010} K. H. Nagai and H. Kori, 
Noise-induced synchronization of a large population of globally coupled nonidentical oscillators, Phys. Rev. E {\bf 81}, 065202(R) (2010).


\bibitem{lai2013} Y. M. Lai and M. A. Porter, Noise-induced synchronization, desynchronization, and clustering in globally coupled nonidentical oscillators, Phys. Rev. E {\bf 88}, 012905 (2013). 


\bibitem{moran1953} P. A. P. Moran, The Statistical Analysis of the Canadian Lynx Cycle. II. Synchronization and meteorology, Aust. J. Zool. 1, 291 (1953).

\bibitem{hudson1999} P. J. Hudson and I. M. Cattadori, The Moran effect: a cause of population synchrony, Trends Ecol. Evol. 14, 1 (1999).

\bibitem{hansen2020} B. B. Hansen, V. Gr{\o}tan, I. Herfindal, and A. M. Lee, The Moran effect revisited: spatial population synchrony under global warming, Ecography 43, 1591 (2020).

\bibitem{hasson2004} U. Hasson, Y. Nir, I. Levy, G. Fuhrmann, and R. Malach, Intersubject Synchronization of Cortical Activity During Natural Vision, Science {\bf 303}, 1634 (2004). 

\bibitem{hasson2009} U. Hasson, R. Malach, and D. J. Heeger, Reliability of cortical activity during natural stimulation, Trends in Cogn. Sci. {\bf 14}, 40 (2009). 

\bibitem{denworth2023} L. Denworth, Brain Waves Synchronize when People Interact, Sci. Am. (July 1, 2023).

\bibitem{lin2009a} K. K. Lin, E. Shea-Brown, and L.-S. Young, 
Spike-time reliability of layered neural oscillator networks,
J. Comput. Neurosci. {\bf 27}, 135 (2009). 

\bibitem{lin2009b} K. K. Lin, E. Shea-Brown, and L.-S. Young, Reliability of layered neural oscillator networks,
Commun. Math. Sci. {\bf 7}, 239 (2009). 

\bibitem{kawamura2008} Y. Kawamura, H. Nakao, K. Arai, H. Kori, and Y. Kuramoto, Collective Phase Sensitivity, Phys. Rev. Lett. {\bf 101}, 024101 (2008).


\bibitem{teramae2009} J. N. Teramae, H. Nakao, and G.B. Ermentrout, Stochastic Phase Reduction for a        General Class of Noisy Limit Cycle Oscillators, Phys. Rev. Lett. {\bf 102}, 194102  (2009).

\bibitem{goldobin2010} D.S. Goldobin, J. N. Teramae, H. Nakao, and G.B. Ermentrout, Dynamics of Limit-Cycle  Oscillators Subject to General Noise, Phys. Rev. Lett. {\bf 105}, 154101 (2010).

\bibitem{canavier2006} C. C. Canavier, Phase response curve, 
Scholarpedia {\bf 1}(12), 1332 (2006).

\bibitem{smeal2010} R. M. Smeal, G. B. Ermentrout, and J. A. White, Phase-response curves and synchronized neural networks, Phil. Trans. R. Soc. B {\bf 365}, 2407 (2010).

\bibitem{higham2001} D. J. Higham, An Algorithmic Introduction to Numerical Simulation of Stochastic Differential Equations, SIAM Review {\bf 43}, 525 (2001).

\bibitem{suli2003} E. S\"{u}li and D. F. Mayers, {\it An Introduction to Numerical Analysis} (Cambridge University Press, Cambridge, 2003).

\bibitem{ott2008} E. Ott and T. M. Antonsen, Low dimensional behavior of large systems of globally coupled oscillators, Chaos {\bf 18}, 037113 (2008).

\bibitem{pimenova2016} A. V. Pimenova, D. S. Goldobin, M. Rosenblum, and A. Pikovsky, Interplay of coupling and common noise at the transition to synchrony in oscillator populations, Sci. Rep. {\bf 6}, 38518 (2016).

\bibitem{goldobin2019} D. S. Goldobin and A.V. Dolmatova, Interplay of the mechanisms of synchronization by common noise and global coupling for a general class of limit-cycle oscillators, Commun. Nonlinear Sci. Numer. Simulat. {\bf 75}, 94 (2019).

\bibitem{bressloff2016} P. C. Bressloff, Stochastic Fokker-Planck equation in random environments, Phys. Rev. E, {\bf 94}, 042129 (2016).

\bibitem{pessoa2014} L. Pessoa, Understanding brain networks and brain organization, Phys. Life Rev. {\bf 11}, 400 (2014).

\end{thebibliography}
\end{document}